\documentclass[english,letterpaper,english,prx,notitlepage,twocolumn,superscriptaddress]{revtex4-2}
\usepackage[utf8]{inputenc}
\usepackage{amsmath}
\usepackage{amssymb}
\usepackage{dsfont}
\usepackage{float}
\usepackage{cancel}
\usepackage{subfiles}
\usepackage[caption=false]{subfig}
\usepackage{stackengine}
\usepackage{hyperref}
\usepackage{nicefrac}

\newcommand{\vect}[1]{\boldsymbol{\mathbf{#1}}}
\newcommand{\Tr}{\operatorname{Tr}}
\newcommand{\hc}{\text{h.c.}}

\newcommand{\acomm}[2]{\left\lbrace #1, #2\right\rbrace}
\newcommand{\lrangle}[1]{\left\langle #1 \right\rangle}
\usepackage{tabularx}
\usepackage{braket}
\newcommand{\spin}[2]{\sigma_{#2}^{\mathrm{#1}}}

\usepackage{color}
\usepackage[usenames,dvipsnames,svgnames,table]{xcolor}
\usepackage{graphicx}
\usepackage{comment}
\usepackage{ulem}
\usepackage{soul}
\definecolor{violet}{rgb}{0.58, 0.0, 0.83}

\begin{document}

\title{Localization dynamics in a centrally coupled system}
\author{Nathan Ng}
\thanks{These authors contributed equally to this work.}
\affiliation{Department of Physics, University of California, Berkeley, CA 94720, USA}
\author{Sebastian Wenderoth}
\thanks{These authors contributed equally to this work.}
\affiliation{Institute of Physics, University of Freiburg, Hermann-Herder-Strasse 3, 79104 Freiburg, Germany}
\author{Rajagopala Reddy Seelam}
\affiliation{Institute of Physics, University of Freiburg, Hermann-Herder-Strasse 3, 79104 Freiburg, Germany}
\author{Eran Rabani}
\affiliation{Department of Chemistry, University of California, Berkeley, CA 94720, USA}
\affiliation{Materials Sciences Division, Lawrence Berkeley National Laboratory, Berkeley, CA 94720, USA}
\affiliation{The Sackler Center for Computational Molecular and Materials Science, Tel Aviv University, Tel Aviv 69978, Israel}
\author{Hans-Dieter Meyer}
\affiliation{Theoretische Chemie, Physikalisch-Chemisches Institut, Universit\"{a}t Heidelberg, INF 229, D-69120 Heidelberg, Germany}
\author{Michael Thoss}
\affiliation{Institute of Physics, University of Freiburg, Hermann-Herder-Strasse 3, 79104 Freiburg, Germany}
\affiliation{EUCOR Centre for Quantum Science and Quantum Computing, University of Freiburg, Hermann-Herder-Strasse 3, 79104 Freiburg, Germany}
\author{Michael Kolodrubetz}
\affiliation{Department of Physics, The University of Texas at Dallas, Richardson, Texas 75080, USA}

\date{\today}

\begin{abstract}
   In systems where interactions couple a central degree of freedom and a bath, one would expect signatures of the bath's phase to be reflected in the dynamics of the central degree of freedom. This has been recently explored in connection with many-body localized baths coupled with a central qubit or a single cavity mode --- systems with growing experimental relevance in various platforms. Such models also have an interesting connection with Floquet many-body localization via quantizing the external drive, although this has been relatively unexplored. Here we adapt the multilayer multiconfigurational time-dependent Hartree (ML-MCTDH) method, a well-known tree tensor network algorithm, to numerically simulate the dynamics of a central degree of freedom, represented by a $d$-level system (qudit), coupled to a disordered interacting 1D spin bath. ML-MCTDH allows us to reach $\approx 10^2$ lattice sites, a far larger system size than what is feasible with exact diagonalization or kernel polynomial methods. From the intermediate time dynamics, we find a well-defined thermodynamic limit for the qudit dynamics upon appropriate rescaling of the system-bath coupling. The spin system shows similar scaling collapse in the Edward-Anderson spin glass order parameter or entanglement entropy at relatively short times. At longer time scales, we see slow growth of the entanglement, which may arise from dephasing mechanisms in the localized system or long-range interactions mediated by the central degree of freedom. Similar signs of localization are shown to appear as well with unscaled system-bath coupling.
\end{abstract}

\maketitle
\section{Introduction}
The advent of controllable quantum simulation platforms allows for novel explorations of quantum coherent phenomena. Certain such architectures have the advantage of using extra degrees of freedom as a way to easily read out properties of a system \cite{Blais2004}. Examples of such setups include cavity QED with ultracold atoms \cite{Davis2019} and superconducting qubit circuits, the latter of which was recently used to simulate the many-body localized (MBL) phase in a 10 qubit chain with long-range interactions mediated by a central resonator \cite{Xu2018}. Given that such platforms are in their early stages, it is important to explore the interplay of disorder-induced localization and mediated long-range interactions, and how they affect the dynamics of localization in these systems.

If localization exists in these systems, it will naturally be many-body localization since the spins hybridize with the central degree of freedom to give non-trivial interactions. Rigorous results on MBL have already been established in one dimensional systems with short ranged interactions \cite{Imbrie2016}. In such a setting, it is a stable phase of matter, with respect to adding short range perturbations, that can coexist with other types of order \cite{Huse2013, Khemani2016}. While strong disorder enables localization, it cannot prevent thermalization if interactions are long-ranged, decaying slower than $r^{-2D}$, where $D$ is the spatial dimension \cite{Yao2014, Maksymov2020}. Even the MBL phase with short-ranged interactions is fragile. It is destroyed upon coupling to a continuum of bath modes \cite{Nandkishore2014} which, intuitively, can provide arbitrary amounts of energy and allow the system to transition between eigenstates of vastly different character. One sees then that there are two ingredients to this delocalization mechanism: a continuum of energies of large enough bandwidth, and hybridization due to effective infinite-ranged interactions mediated by the non-Markovian bath. 

In fact, for a specific type of memoryless bath, nonergodicity does survive. This is the case of Floquet MBL, in which an MBL system is subjected to an external periodic drive with frequency $\Omega$ modeled as a time-dependent Hamiltonian acting on the system \cite{Abanin2016,Ponte2015prl}. The failure of thermalization is due to the the inability of the system to absorb energy in quanta of $\hbar \Omega$, which itself is a consequence of the discreteness of the energy spectrum. The external drive, however, is not inherently dynamical and thus does not capture the backaction present in a fully quantum mechanical system.

In this work, we consider the time evolution of such a system obtained by treating the Floquet drive as a quantum degree of freedom. Specifically, we consider a localized system globally coupled to a $d$-level system (qudit) with finite energy spacing, similar to \cite{Ng2019}. When the qudit is a two-level system, it was shown that localization does not survive at any finite coupling \cite{Ponte2017, Hetterich2018}. But when it is instead a $d>2$ level system, localization was argued to survive under certain conditions \cite{Ng2019}. It is not known, however, what dynamical signatures should be expected in such regimes since the geometry and spin-spin interactions in the system limits the efficiency of usual computational approaches using matrix product operators. We bridge this gap by numerically simulating the non-equilibrium dynamics at much larger system sizes than previously considered. This is done using the multilayer-multiconfigurational time-dependent Hartree (ML-MCTDH) method, which solves the Schr\"odinger equation using the time-dependent variational principle on the manifold of wavefunctions represented by certain tree tensor networks \cite{Wang2003,Manthe2008,MCTDHBook,Oriol2011,Wang2015}.

We furthermore explore the possibility that the additional degree of freedom can provide alternative, nondestructive diagnostics of localization. In experimental settings, the usual observables signaling nonergodic behavior are correlation functions such as the occupation imbalance between odd and even sites of the lattice \cite{Schreiber2015}. More sophisticated setups may attempt to perform tomographic measurements to reconstruct the reduced density matrix for a subsystem and show logarithmic growth of entanglement entropy \cite{Xu2018}, or to measure the energy spectrum of the system in order to retrieve energy level spacing statistics \cite{Roushan2017}. Though these metrics serve as gold standards in characterizing MBL, the latter two methods are difficult to scale with larger systems. In our model, since quantum fluctuations of the spins necessarily involve the qudit, there may be signatures of (de)localization imprinted into the qudit dynamics. Such a possibility has been explored in autocorrelations of qudit observables \cite{Hetterich2018} probing the energy level statistics, as well as dynamics of the occupation number \cite{Sierant2019} by measuring the light intensity output by a single mode cavity. In this work we show that the qudit qualitatively changes the spin chain dynamics, and elucidate the timescale on which this occurs. This provides some insight into the breakdown of localization, and the possible role that non-Floquet physics may play in it.

The structure of this paper is as follows: we will discuss our model and its localization in connection to Floquet MBL; review the essentials of ML-MCTDH, which we then apply to study intermediate time dynamics; present results on thermalizing and nonthermalizing behaviors in dynamical metrics; and discuss what may be expected in experiments, where control over the central coupling may be limited in range.

\section{Model}
\label{sec:model}
We consider a simplified model of many-body localization by coupling a one-dimensional  chain of qubits (spins-$1/2$) via global interactions with a central qudit:
\begin{align}
	\label{eq:model}
	H &= H_0 + \Omega \hat{\tau}^z + \gamma H_1 \left( \hat{\tau}^+ + \hc \right), \\
	H_0 &= \sum_{i=1}^L h \xi_i \spin{z}{i} + g \spin{z}{i} \spin{z}{i+1}, \qquad H_1 = \sum_{i=1}^L \spin{x}{i}, 
\nonumber
\end{align}
where $\hat{\tau}^z = \sum_{n=1}^d n |n\rangle\langle n|$, $\hat{\tau}^+ = \sum_{n=1}^{d-1} |n+1\rangle\langle n|$, and the operators $H_0$ and $H_1$ act only on the spin subspace. The states $|n\rangle$ label the states of the central qudit. Here, $h = 1.3$, $g = 1.07$, $\Omega = \pi/0.8$, and $\xi_i$ is a random variable drawn uniformly from $(-1,1)$. When the model with these parameters is mapped on to the corresponding Floquet system (i.e., $d\to\infty$), it shows a localization-delocalization transition at a critical coupling $\gamma_c \lesssim 0.3$ \footnote{This estimate is subject to strong finite size effects}. We restrict our discussion to $\gamma$ either deep in the localized phase ($\gamma < 0.2$) or deep in the ergodic phase ($\gamma\approx 1$). Finally, throughout this paper we restrict ourselves to central qudit size $d=7$, which is large enough to display Floquet-like behavior but small enough that the finite qudit size plays an important role.

The spin part of the Hamiltonian, $H_0$, is a trivial antiferromagnetic Ising chain with longitudinal on-site disorder. The diagonal nature of $H_0$ in the $z$-basis yields trivial localization in the eigenstates. This manifests in eigenstates $|\psi_n\rangle$ as vanishing site-averaged magnetization $L^{-1}\sum_i \lrangle{\psi_n | \spin{z}{i} | \psi_n}$ and maximal value of the spin-glass parameter, $q = L^{-1}\sum_i \lrangle{\psi_n | \spin{z}{i} | \psi_n}^2 = 1$ at high energy densities. Values of $q\approx 1$ suggest that the eigenstates are described mostly by a single pattern of magnetization. Introducing a small coupling to the qudit without longitudinal disorder induces hybridizations that push $q\to 0$. We find that it is necessary to have both qudit coupling and strong disorder to preserve the nonergodicity when probing the system in the middle of the many-body spectrum, where the density of states (DOS) is the greatest.

Several features distinguish our model from those studied previously. While Nandkishore et al.\ \cite{Nandkishore2014} coupled a ``fully MBL'' system to an interacting bath of bosons, the qudit we present here is not bath-like because it does not have a continuous DOS. The model of thermal inclusions studied by Ponte et al.\ \cite{Ponte2017} closely resembles ours, but crucially we place a constant ``magnetic field'' $\Omega \hat{\tau}_z$ on the qudit, thus selecting a preferred direction for the central spin. This greatly impacts the ease with which the qudit fluctuates, which in turn can regulate transitions in the spin states leading to delocalization. 

Recent studies have examined how localization can persist in the presence of long-ranged interactions \cite{Sierant2019, Maksymov2020} or with central coupling to a single degree of freedom yielding an effective Hamiltonian with long-ranged interactions \cite{Ponte2017, Hetterich2018}. With the exception of a numerical study \cite{Sierant2019}, these past works have noted that preservation of localization in the thermodynamic limit requires increasing the disorder strength with increasing system size or decreasing the strength of central coupling as $\gamma\to \gamma/L$. Reducing the coupling strength in this way renders the long-ranged part of the effective Hamiltonian for the spin chain subextensive. This is also reflected in the dynamics of the qudit as its transition rate vanishes.

On the other hand, the existence of Floquet MBL affords a different pathway to the coexistence of localization and central coupling. In that context, the persistence of MBL is not due to a vanishing coupling to the external drive, but to a suppression of mixing between different localized eigenstates of the undriven system. This picture suggests that an effective Hamiltonian for only the spin degrees of freedom should show localized behavior. This is indeed the case, as previous work based on the high frequency expansion has shown \cite{Ng2019}. In this limit of $\Omega \to \infty$ the spins are governed by an effective Hamiltonian diagonal in the qudit basis, reproducing the eigenenergies modulo an integer multiple of $\Omega$:
\begin{align}
\label{eq:HFE}
    H_{\text{eff}} = H_0 + (H_1)^2 \frac{|d\rangle\langle d| - |1\rangle\langle 1|}{\Omega} + O(\Omega^{-2}).
\end{align}At lowest order in $\Omega^{-1}$, we see that possible delocalization is reserved only for states with $|1\rangle$ or $|d\rangle$, as $(H_1)^2$ induces all-to-all coupling. Increasing $L$ without increasing $d$, as we do in this paper, means that eigenstates occupying $|1\rangle$ will eventually encroach upon the middle of the spectrum and contribute to the quench dynamics we study. This can be seen from the density of states when $H_0$ is dominant as it follows $\rho(E) \propto \exp \left( - \frac{E^2}{(J \sqrt{L})^2} \right)$ for energy scale $J\sim O(1)$, meaning $\rho(E)$ will grow wider with increasing $L$. An energetically dominant $(H_1)^2$ term will both delocalize the eigenstates and deform the Gaussian density of states in the thermodynamic limit.

\begin{figure}[h]
  \includegraphics[width=0.95\columnwidth]{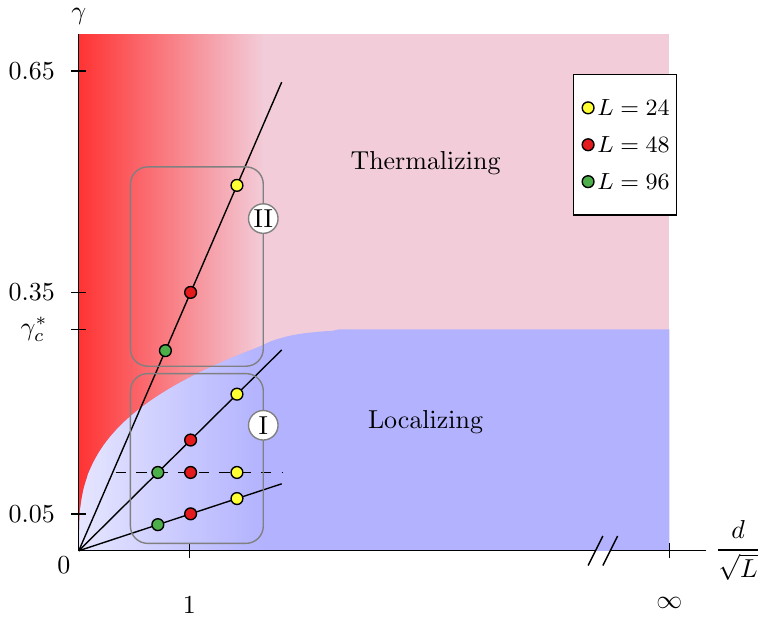}
  \caption{Schematic phase diagram for the coupled system \eqref{eq:model}, along with the parameters for which we present numerical results from ML-MCTDH. The rough phase boundaries are determined from numerics and analytical arguments. The data are separated into three solid segments -- the strong, intermediate and weak couplings from top to bottom. The angle of the segments comes from fixing the qudit size to $d=7$ and scaling the central coupling $\gamma \propto L^{-1/2}$. For ease of discussion, we group the three coupling regimes as region I (weak and intermediate) and region II (strong).}
  \label{fig:phaseDiagram}
\end{figure}

We thus assume $\gamma$ to be small enough such that neither outcome occurs, and ask when this picture will naively break down. In such a limit, we can treat the $H_1^2$ field term in a mean field fashion for each eigenstate:
\[ H_{\text{eff}} \approx H_0 + \frac{\gamma^2 L}{\Omega} + \sum_i \frac{\gamma^2}{\Omega}\lrangle{\sum_{j\neq i} \spin{x}{j}} \spin{x}{i}, \] where the effective field $\lrangle{\sum_{j\neq i} \spin{x}{j}}$ in an eigenstate must be determined self-consistently. For a typical eigenstate, this field should have value $\sim f(\gamma) \sqrt{L}$, where $f(\gamma)$ must vanish when $\gamma=0$. This is the case when $\sum_j \lrangle{\spin{x}{j}}$ is the sum of $L-1$ independent random variables, and the finite $\gamma$ eigenstates are assumed to be perturbatively connected to a corresponding $\gamma=0$ eigenstate. For this model, we take the lowest order approximation $f(\gamma) \approx f_1 \gamma$. With this assumption \footnote{See Supplemental Material at URL for discussion on the cumulant expansion, an ansatz for the steady state of the thermalizing phase, and perturbation theory using multiple scales.}, the effective transverse field on site $i$ will begin to compete with the longitudinal fields in $H_0$ when $\gamma^2 \lrangle{\sum_{j\neq i} \spin{x}{j}} \sim O(g, h_i) \sim O(1)$. For the high energy density eigenstates we are interested in, this effective field will inhibit spin glass ordering and the system should obey the eigenstate thermalization hypothesis. Thus, $\gamma \propto L^{-1/6}$ should serve as a rough separatrix between thermalizing and athermal behaviors. Furthermore, couplings that tend to zero faster than $L^{-1/6}$ will realize a trivial limit, where the localization comes entirely from $H_0$. 
The region where this is argument expected to be most significant is denoted in Fig.\ \ref{fig:phaseDiagram} through a color gradient starting around $d/\sqrt{L} \sim 1$. 
Note that, in general models where $H_1$ includes operators diagonal in the z-basis, we would have $f(0) \neq 0$; in this case the scaling is replaced by $\gamma \sim L^{-1/4}$.

Besides scaling the coupling to zero, the all-to-all interactions can be avoided by ensuring that eigenstates occupying levels $|d\rangle$ or $|1\rangle$ in the qudit do not participate in the dynamics. For quenches starting from the middle of the many-body spectrum, this condition can be ensured by keeping the qudit size $d$ sufficiently large compared to the typical width of the 1D many-body density of states, $\sqrt{L}$. Dynamics in this limit should closely resemble Floquet physics, since the fluctuations producing effective long-ranged interactions will cancel out after accounting for the processes in which the intermediate qudit state changes by $+1$ or $-1$. Away from this limit, when $d/\sqrt{L} \lesssim O(1)$, the all-to-all interactions are unavoidable. The threshold value of $d/\sqrt{L}$ for delocalization should decrease as the coupling is decreased. These arguments are summarized schematically in Fig.\ \ref{fig:phaseDiagram}.

\begin{figure*}
\begin{minipage}{0.45\textwidth}
\begin{align*}
\ket{\Psi} &= \sum_{j_1=1}^{N_{1}}... \sum_{j_P=1}^{N_{1}} A_{j_1,...,j_P}(t) \prod_{\kappa=1}^P \ket{\varphi_{j_\kappa}^{(\kappa)}(t)}, \\
\ket{\varphi_{j_\kappa}^{(\kappa)}(t)} &= \sum_{i_1=1}^{N_{2}}... \sum_{i_{Q(\kappa)}=1}^{N_{2}} B_{i_1,...,i_{Q(\kappa)}}^{\kappa, j_\kappa}(t) \prod_{q=1}^{Q(\kappa)} \ket{\nu_{i_q}^{(\kappa,q)}(t)}, \\
&..., \nonumber
\end{align*}
\end{minipage}
\begin{minipage}{0.45\textwidth}
\includegraphics[scale=0.6]{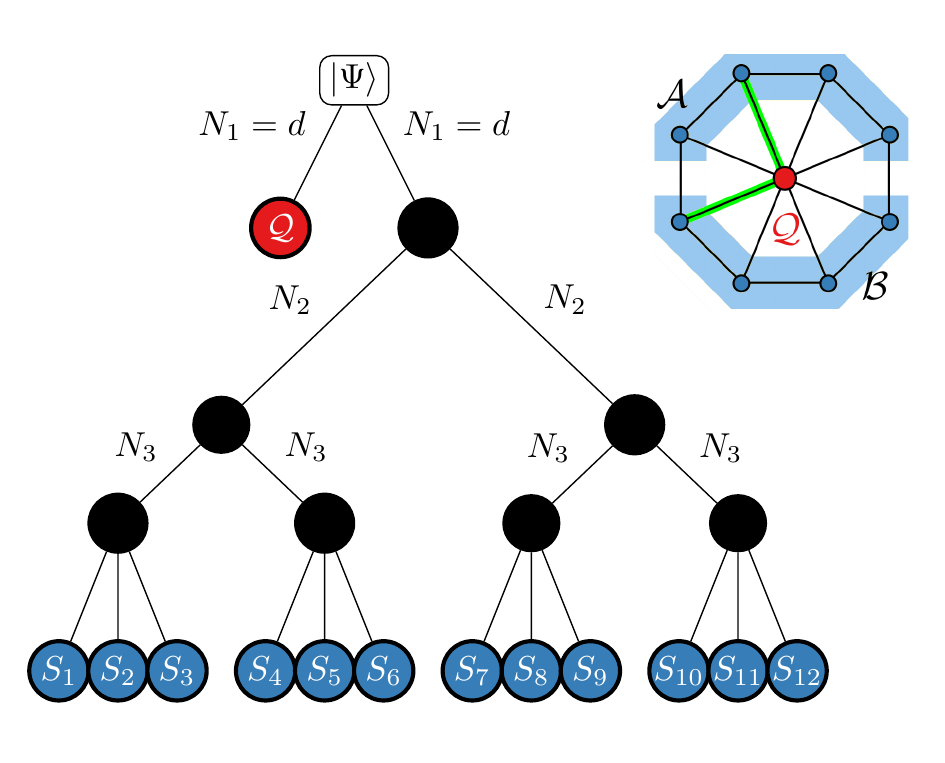}
\end{minipage}
\caption{Expansion of the wave function $\ket{\Psi}$ and the first layer single-particle functions $\ket{\varphi_{j_\kappa}^{(\kappa)}(t)}$ used in the ML-MCTDH approach \textbf{(left)} and a schematic representation of the tree structure of the wave function \textbf{(right)}. The black dots represent single-particle functions (SPFs). The red dot represents the qudit degree of freedom and the blue dots represent the spin degrees of freedom. The binary expansion of the spin wave function is symmetric and, thus, we choose the numbers of SPFs within one layer to be equal. In the example shown, only three spins are grouped together in the lowest layer for better visualization. In the calculation, however, groups of up to 12 spins in the lowest layer are used.}
\label{fig:mctdh_tree}
\end{figure*}

There are two important ways to think of this system and its dynamics: either as a combined many-body systems with localized and delocalized phases, as was done in the previous paragraph, or as a central qudit interacting with an unusual, localized, spin bath. From this latter viewpoint, it will be useful to consider scaling the system bath coupling $\gamma \sim 1/\sqrt{L}$, since that will be shown to achieve a well-defined thermodynamic ($L\to\infty$) limit. This scaled coupling will be used in the majority of our simulations, and is covered in more detail in Section \ref{sec:scaledGamma}. For now, we note that $\gamma \sim L^{-1/2}$ scales to zero faster than the $L^{-1/6}$ that we predict is required for MBL. Therefore, at sufficiently late times, we predict MBL with our scaled coupling.

In this model we use qudits for numerical simplicity due to their finite Hilbert spaces. However, our conclusions can be easily applied also to the case where the central degree of freedom is a single bosonic mode, such as in cavity QED or superconducting circuits. In these setups we expect similar dynamical behaviors when the central coupling is appropriately scaled \cite{Ng2019}.

\section{Numerical method}
\label{sec:numerics}
\begin{figure*}
  \centerline{\includegraphics[width=0.99\textwidth, trim={0 0 0 0}, clip]{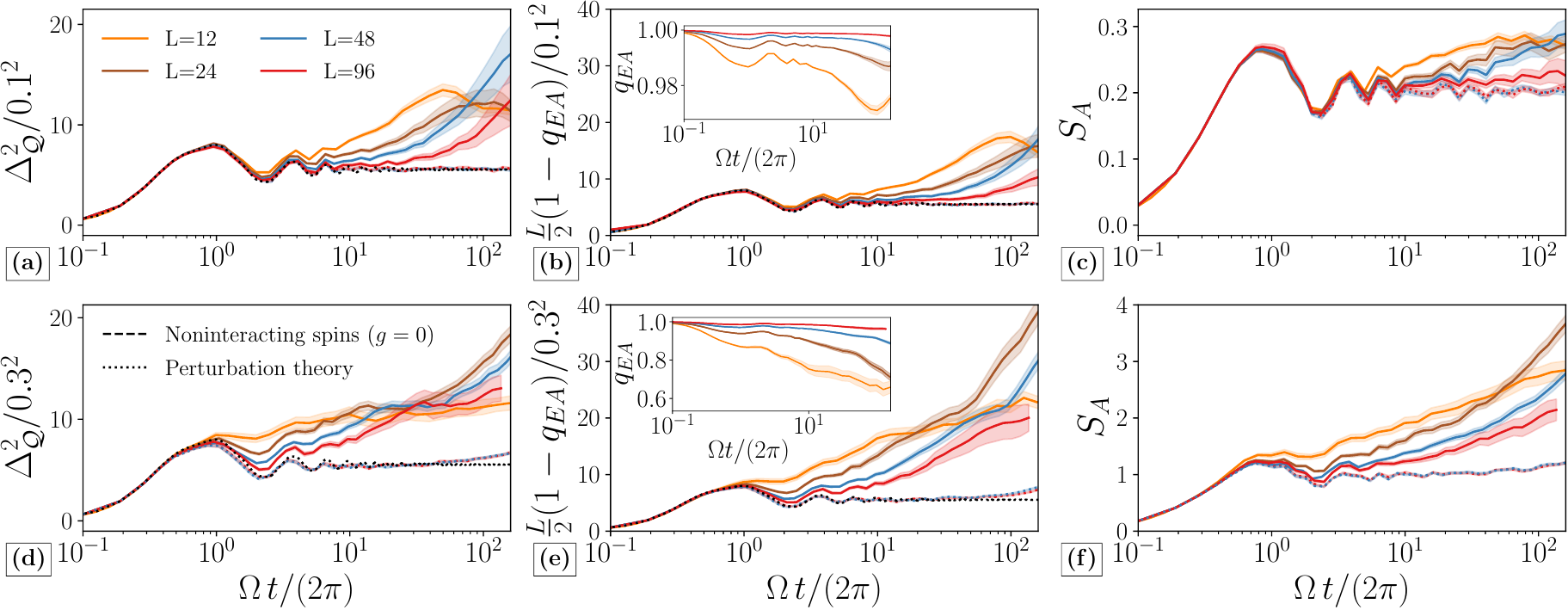}}
  \caption{Dynamics from an initial ``super-Neel'' state in the regimes of weak (top row, $\gamma =0.1/\sqrt{L/12}$) and intermediate coupling (bottom row, $\gamma =0.3/\sqrt{L/12}$). The data is averaged over $O(10^2)$ -- $O(10^3)$ disorder realizations, with the shaded bands indicating deviations of $\pm 1$ standard error of the mean. 
  \textbf{(left)} Variance of qudit occupations $\Delta^2_Q = \lrangle{(\tau^z)^2} - \lrangle{\tau^z}^2$, \textbf{(center)} deviation from perfect spin glass order $1-q_{EA}$, and \textbf{(right)} entanglement entropy $S_A$ between a contiguous half of the spin chain and its complement. The observables $\Delta_{\mathcal{Q}}^2$ and $1-q_{\text{EA}}$ have been appropriately rescaled \cite{Note2} to show their coincidence (except a factor of 2) at early times for weak coupling. The dynamics are observed to converge to a single curve (black dotted line) and appear to be consistent with the dynamics without the nearest neighbor Ising coupling (dashed lines) as $L\to\infty$.}
  \label{fig:shortTime}
\end{figure*}

The non-local interaction induced by the centrally coupled qudit makes the simulation based on matrix product operator techniques like time-evolving block decimation \cite{Paeckel2019} inefficient. And while alternative approaches such as the Floquet-Keldysh DMFT \cite{Lubatsch2019} exist, they are valid only in the well-studied Floquet limit in which the interesting mediated all-to-all couplings are negligible. Thus, we instead employ the Multilayer Multiconfiguration Time-Dependent Hartree (ML-MCTDH) method \cite{Wang2003,Manthe2008,Oriol2011,Wang2015, HeidelbergMCTDH} which has been used to study similar systems in the past, e.g.\ a two-level system coupled to a bath of noninteracting spins \cite{Wang2012}. The ML-MCTDH method generalizes the original MCTDH method \cite{Meyer1990,man92:3199,Beck2000,MCTDHBook,mey12:351} for applications to significantly larger systems. The ML-MCTDH approach represents a rigorous variational basis-set method, which uses a multiconfiguration expansion of the wave function, employing time-dependent basis functions and a hierarchical multilayer representation. Within this framework the wave function is recursively expanded as a superposition of Hartree products as depicted in Fig. \ref{fig:mctdh_tree}. Here, $\ket{\varphi_{j_\kappa}^{(\kappa)}(t)}$, $\ket{\nu_{i_q}^{(\kappa,q)}(t)},\ldots,$ are the so-called ``single-particle functions'' (SPFs) for the first, second, etc.\ layer and the coefficients $A_{j_1,...,j_N}$, $B_{i_1,...,i_{Q(\kappa)}}^{\kappa, j_\kappa}$ are the expansion coefficients of the first, second, etc.\ layer. Despite their name, the SPFs describe multiple degrees of freedom, see Fig.\ \ref{fig:mctdh_tree}. The ML-MCTDH equations of motions for the expansion coefficients and the single-particle functions are obtained by applying the Dirac-Frenkel variational principle \cite{Wang2003, Wang2009}, thus ensuring convergence to the solution of the time-dependent Schrödinger equation upon increasing the number of SPFs. In principle, the recursive multilayer expansion, which corresponds to a hierarchical tensor decomposition in the form of a tensor tree network, can be carried out to an arbitrary number of layers. In practice, the multilayer hierarchy is terminated at a particular level by expanding the single-particle functions in the deepest layer in terms of time-independent basis functions.

In the present application of the ML-MCTDH method, we separate the qudit wave function and the spin chain wave function in the uppermost layer as depicted schematically in Fig.\ \ref{fig:mctdh_tree}. The wave function of the spin chain is then further expanded in a binary tree (i.e.\ $P=Q=2$) up to the lowest layer, which comprises blocks of up to 12 spins. Each of the lowest blocks is expanded in the time-independent local basis of the underlying Hilbert space. Regarding the number of SPFs in the first layer, $N_1$, it can be shown that $N_1>d$ leads to redundant configurations in the expansion \cite{Meyer1990}, and thus, we set $N_1=d$ in all calculations. The required number of SPFs in the other layers of the expansion of the spin-chain wave function was determined by thorough convergence tests and depends on the coupling strength $\gamma$. In general, fewer SPFs are needed for smaller coupling strengths. For L=24, two dynamical layers are employed and the required number $N_2$ of SPFs varies from 30 to 120 SPFs. For L=48, a three layer scheme is used where the number of SPFs in the lowest layer varies from 10 to 30 and in the highest layer from 20 to 60 SPFs. For L=96, four layers are employed with SPFs which  vary from 10 to 20 in the lowest and from 35 to 50 in the highest layer.

\section{Results for scaled coupling}
\label{sec:scaledGamma}

We examine the system at infinite temperature by focusing on states in the middle of the many-body spectrum, which have energies close to the midpoint between the maximal and minimal energies of the coupled system, $(E_{\text{max}}$ and $E_{\text{min}})$ respectively. We take $\gamma=0$ for $t<0$ with the spins in a ``super-Neel'' state  $|\downarrow \downarrow \uparrow \uparrow \ldots \rangle$ and the qudit occupying its middle state $|(d+1)/2\rangle$. The coupling is switched on instantaneously at $t=0$ to a finite value. The super-Neel state is on average a zero energy eigenstate of $H_0$ and has subextensive energy variance, making it a suitable microcanonical probe. Thus, when the system is thermalizing and shows ensemble equivalence, we expect similar dynamics compared to ones obtained through averaging over random initial product states, mimicking an infinite temperature canonical ensemble.

As there are different dynamical behaviors in our model, we shall organize our discussion around the schematic phase diagram in Fig.\ \ref{fig:phaseDiagram}, similar to the one first introduced in \cite{Ng2019}. In this first section, we will consider scaling the coupling as $\gamma \sim 1/\sqrt{L}$, corresponding to the three solid lines in the phase diagram which, from top to bottom, will be referred to as the strong, intermediate, and weak coupling regimes. The orientation of these cuts comes from the $1/\sqrt{L}$ scaling of $\gamma$. This is natural if we think of the qudit as our main object of interest, as it gives a well-defined thermodynamic limit for the qudit when it is coupled to a non-interacting bath, such as in the spin-boson model \cite{Weiss2012}. This scaling reproduces the Kac prescription \cite{Kac1963} for the all-to-all term in the effective Hamiltonian ensuring also the existence of a thermodynamic limit for the spins. Specifically, we scale $\gamma$ using the following formula:
\begin{equation}
\gamma  = \gamma_0 \sqrt{\frac{L_0}{L}}
\label{eq:gamma_scaling}
\end{equation}
where $L_0=12$ throughout for convenience, such that $\gamma=\gamma_0$ at $L=12$. $\gamma_0$ sets the overall strength of the coupling. We will consider three regimes, indicated by the solid lines in Fig.\ \ref{fig:phaseDiagram}: weak coupling ($\gamma_0=0.1$), intermediate coupling ($\gamma_0=0.3$), and strong coupling ($\gamma_0=0.7$).

\begin{figure*}
  \centerline{\includegraphics[width=0.99\textwidth, trim={0 0 0 0}, clip]{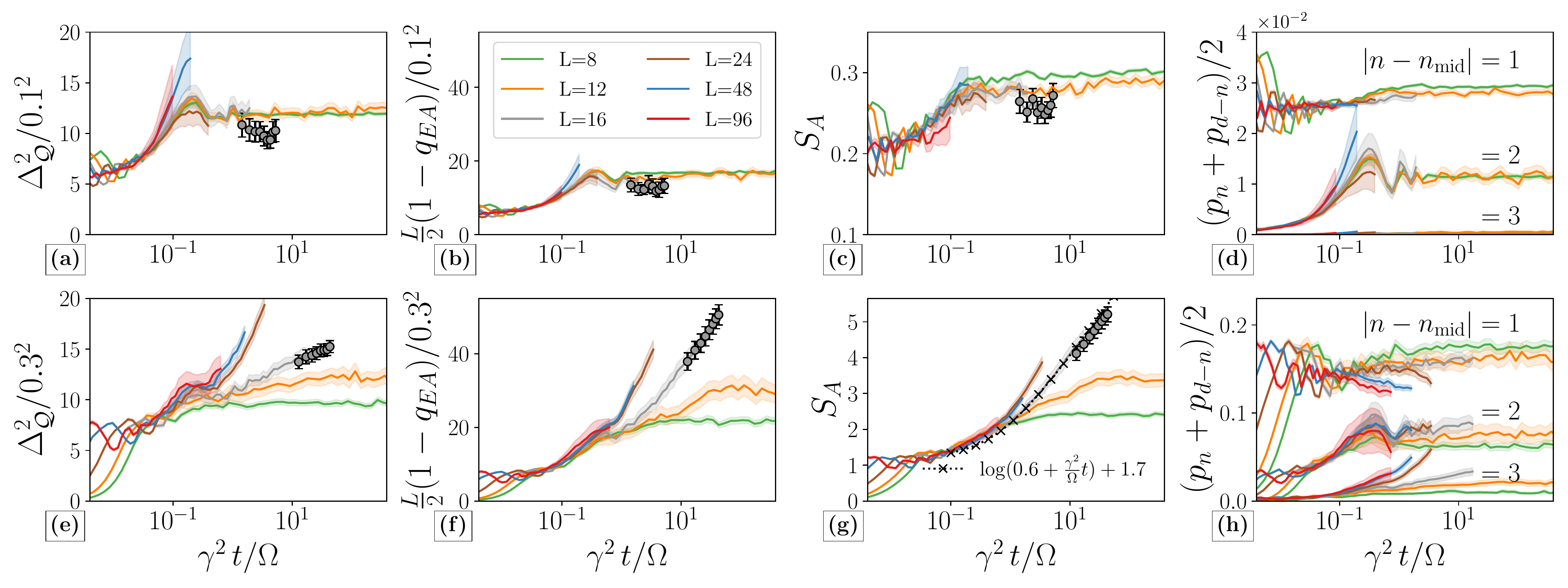}}
  \caption{Same as in Fig.\ \ref{fig:shortTime}, but with time rescaled by the system size-dependent coupling $\gamma $. Grey dots are independent calculations using the kernel polynomial method, aimed to extend the maximum time from $\Omega t/(2\pi) \sim 6\times 10^2$ to $\Omega t/(2\pi) \sim 1.9 \times 10^3$. Between the weak \textbf{(a,b,c,d)} and intermediate \textbf{(e,f,g,h)} coupling regimes, there is a qualitative shift in the long time behavior of both the qudit- and spin-only observables. This data is suggestive of logarithmic growth in the entanglement entropy becoming the dominant characteristic after $t \sim 1/\gamma ^2$. \textbf{(d,h)} Occupations $p_n$ of the qudit levels, symmetrized around the middle level $|n_{\text{mid}}\rangle = |4\rangle$. While both the weak \textbf{(d)} and intermediate \textbf{(h)} coupling regimes have most of their populations concentrated the initial occupied level, $|4\rangle$, the latter case has a much greater fraction of the total population in the extremes of the qudit's states. The values of $p_n$ for $|n-n_{\text{mid}}|=3$ in the left panel are too small ($\sim O(10^{-4}-10^{-3})$) for the scale.} 
  \label{fig:intTime}
\end{figure*}

\subsection{Weak and intermediate coupling (region I)}
\label{weakerCoupling}

The first cases we consider are weak and intermediate coupling, which are labeled Region I in Fig.\ \ref{fig:phaseDiagram}. These are both at sufficiently small $\gamma_0$ that we expect MBL for the largest accessible system sizes, but for intermediate coupling ($\gamma_0 = 0.3$) the system will be near the phase transition for small $L$. Three observables -- the qudit variance $\Delta_Q^2$ (Eq. \ref{eq:Delta_Q_def}), the spin glass order parameter $q_\mathrm{EA}$ (Eq. \ref{eq:q_EA_definition}), and the entanglement entropy of the half chain $S_A$ (Eq. \ref{eq:ent_entr_def}) -- are plotted in Figs.\ \ref{fig:shortTime} and \ref{fig:intTime}, which correspond to identical data with different scaling of the time axis. The origin of this scaling will be clarified shortly.

Note first that, by preparing both the qudit and the spins in highly excited states, one would normally expect the system to relax quickly to a featureless ``infinite temperature'' equilibrium. That is, all internal levels of the qudit should be equally occupied, and the spins should be paramagnetic and translationally invariant. This is not true for the disordered system we study, as the numerics demonstrate in Fig.\ \ref{fig:shortTime}: for sufficiently small coupling, the system shows localization in both the qudit and its the surrounding spins. The former is signaled by the variance of the qudit occupations 
\begin{equation}
\Delta^2_{\mathcal{Q}} \equiv \lrangle{(\hat{\tau}^z)^2} - \lrangle{\hat{\tau}^z}^2,    
\label{eq:Delta_Q_def}
\end{equation}
which saturates to a quantity far below that of the uniform limit, $\Delta^2_{\mathcal{Q}} = (d^2-1)/12 = 4$. Furthermore, the different system sizes exhibit scaling collapse of $\Delta^2_{\mathcal{Q}}$ up to a time scale $t \sim 1/\gamma $. This is a property of the scaled $\gamma$, as it implies that the spin chain acts as a bath for the qudit with a well-defined thermodynamic limit. More specifically, it can be shown that the considered model with scaled coupling $\gamma \propto L^{-1/2}$ fulfills linear response in the thermodynamic limit, meaning that the effect of the spin environment on the qudit is captured by the first two cumulants of the influence functional \cite{Feynman1963,Makri1999,Wang07}. For our model, the first cumulant vanishes and thus the reduced qudit dynamics is determined by the second cumulant, given by the force-force autocorrelation function of the spin chain. This also means that one can construct an effective harmonic bath whose correlation function is the same as that of the spin chain resulting in the same reduced qudit dynamics \cite{Makri1999}. For our model, the effective harmonic bath is characterized by a spectral density which depends in general on the initial state, the random local fields and the spin-spin coupling $g$. For the specific initial state considered here, the spectral density of the effective harmonic bath is equal to the probability distribution of twice the random local fields, and thus is independent of $g$.

Having established scaling collapse of the qudit variance, we now turn our attention to dynamics of the spin chain, starting with the spin glass order parameter 
\begin{equation}
q_{\text{EA}}(t) \equiv L^{-1}\sum_i\lrangle{\psi|\spin{z}{i}(t)\spin{z}{i}(0)|\psi}.
\label{eq:q_EA_definition}
\end{equation}
Unlike the qudit variance, the spin glass order parameter displays marked drifts with system size (see insets of Fig.\ \ref{fig:shortTime}(b,e)). The tendency of $q_{\text{EA}}(t) \to 1$ comes from our choice of scaling $\gamma$, since $\gamma$ controls the strength of a local transverse field and thus governs the rate and magnitude of a single spin's precession. On reachable timescales $t\lesssim 10^2$, the largest system size $L=96$ has near perfect memory of the initial state. This behavior is consistent with our claim that the scaling of $\gamma \sim 1/\sqrt{L}$ towards zero with increased system size is sufficiently fast that the system will flow to MBL for arbitrary $\gamma_0$, although proving MBL would require evolution to much later times than we can access. 

Though the usefulness of the influence functional approach is restricted to the qudit, we should -- by virtue of the fact that the initial spin dynamics are driven by interactions with the qudit (for initial product states like the super-Neel state we have chosen) -- find that the spin observables are linked to the qudit's. The spin observables should therefore enjoy a similar limiting behavior as $\gamma  \propto L^{-1/2} \to 0$. We indeed show this to be the case within first order perturbation theory. In \cite{Note2}, we perform time-dependent perturbation theory using the method of multiple scales. We solve for the time evolution operator perturbatively by introducing new ``independent'' timescales $t$, $t' \equiv \gamma t$, $t'' \equiv \gamma^2 t$, $\ldots$, which allow for control over secular terms growing with $t$. In the thermodynamic limit with scaled coupling, we find that the dynamics of the qudit are described perturbatively to first order up to time $O(1/\gamma )$ (dotted lines in Fig.\ \ref{fig:shortTime}), providing a complementary approach to the linear response solution from the influence functional formalism. The perturbative calculation also demonstrates that spin observables should exhibit similar gradual convergence to a single limit up to timescales $t\sim O(1/\gamma )$. Remarkably, the connection between qudit variance and the spin glass order parameter is even more precise in this limit; they collapse to a single, universal curve in the thermodynamic limit upon scaling as $\Delta_\mathrm{Q}^2/\gamma_0^2$ and $(1-q_\mathrm{EA})L/(2\gamma_0^2)$, as seen in Fig.\ \ref{fig:shortTime}(a,b,d,e). Physically, this comes from the fact that a single perturbative excitation of the qudit through the $\hat{\tau}^+ + \hat{\tau}^-$ component of $H_1$ gives a single spin flip excitation of the spin chain through $\spin{x}{j}$.

Finally, we consider the entanglement entropy
\begin{equation}
    S_A = -\mathrm{Tr}\left[\rho_A \log_2 \rho_A\right]
    \label{eq:ent_entr_def}
\end{equation}
between a contiguous half of the spins with the rest of the system, which is a defining feature in many body localization. Here $\rho_A$ is the reduced density matrix of half of the spin system, e.g., sites $1$ through $L/2$. As with the previous two quantities, there appears to be a gradual convergence of $S_A$  to a universal curve with increasing $L$, although unlike the other observables, the entanglement depends on the strength of the coupling prefactor $\gamma_0$. By turning off the Ising interaction $g$ (dashed lines in Fig.\ \ref{fig:shortTime}), we see that the dynamics of entanglement at short times $ \lesssim O(1)$ are unchanged -- as predicted from time-dependent perturbation theory -- while growth of entanglement at intermediate times is dependent on this $\spin{z}{i}\spin{z}{i+1}$ interaction. 

These observations about the short-time dynamics hold for both weak and intermediate coupling, as seen in Fig.\ \ref{fig:shortTime}. However, we can identify a slower timescale beyond $t \lesssim O(1/\gamma )$ from first order perturbation theory, on which the Ising interactions start to play a role. In Fig.\ \ref{fig:intTime}, the same data is plotted upon rescaling the time by $t/\gamma ^{-2}$. The observables are seen to roughly collapse for both the weak and intermediate couplings and, for intermediate couplings, entanglement in particular shows interesting intermediate time behavior. While the collapse is imperfect, we note a few salient features. First, deep in the localized (weak coupling) regime, the spread of the qudit occupation, the growth of bipartite entanglement entropy, and the decay of the spin-glass order parameter appear to be arrested at long times. It is unclear whether the observables will continue to grow at later times, but our data leaves open the possibility that they saturate and that the asymptotic value may be system-size independent under the chosen scaling. Second, the dynamics of the qudit appear to be correlated with dynamics of the spins, albeit with a slight time delay. Finally, in the intermediate coupling regime, the entanglement entropy continues to grow at late times. For $L=16$, there appears to be a logarithmic growth over three decades in rescaled time (see Fig.\ \ref{fig:intTime}g). The same may be true for the $L\geq 24$, but we have insufficient data to decisively prove slow growth over several decades. As seen in Fig.\ \ref{fig:intTime}h, the period of potentially logarithmic growth coexists with the period of finite occupation in the edge of the qudit spectrum (states $n=1$ and $d$), for which the high-frequency expansion yields all-to-all interactions (see Eq. \ref{eq:HFE}).

It is unclear what drives the logarithmic behavior. When focusing on the bipartite entanglement entropy, two generic mechanisms have been studied in recent years: the slow dephasing from a quench due to interactions between exponentially localized (quasi-local) operators \cite{Znidaric2008,Abanin2019}, and the linearly diverging semiclassical trajectories of the collective spin state \cite{Lerose2020} in long ranged interacting spin systems.
In the former case, it has been found that the slope of the logarithmic growth is independent of the strength of interactions \cite{Kjall2014}. This does not appear to be the case in our numerics, with the larger system sizes $L\geq 24$ ostensibly displaying log growth with a larger prefactor than in the $L=16$ case. Moreover, there does not appear to be any logarithmic trend when the system is deep in the localized phase (see top row of Fig.\ \ref{fig:intTime}). If conserved quasi-local operators do exist in this system, then our results would suggest that their localization lengths are strongly dependent on the coupling $\gamma$. 

Another possibility for the appearance of logarithmic growth of $S_A$ could come from the mediated all-to-all interactions predicted in the effective Hamiltonian (Eq. \ref{eq:HFE}). In our qudit system, long ranged interactions begin to play a significant role when the extremal states of the qudit are occupied (see discussion in Sec.\ \ref{sec:model}). It was argued that these mediated interactions are responsible for the localization-delocalization transition upon decreasing $d/\sqrt{L}$, shown in Fig.\ \ref{fig:phaseDiagram}. Consistent with this, we see significantly greater occupation in the extremal qudit states for intermediate couplings -- where logarithmic growth is seen -- compared to weak couplings (see Fig.\ \ref{fig:intTime}(d,h)). It is also clear that the slow growth of $\Delta^2_{\mathcal{Q}}$ for intermediate couplings is due in part to the slow growth in the occupations of the $|1\rangle$ and $|7\rangle$ states. 

Regardless of the origin of slow growth, finite occupation at the extremes of the qudit spectrum implies a departure from the Floquet regime. Our finite time numerics are unable to resolve whether this implies delocalization. Should this mechanism give rise to a sharp localization transition, it would possibly be of a different character from the extensively studied MBL transition based on ergodic grains thermalizing nearby insulating regions through short range interactions \cite{Vosk2013, Dumitrescu2017, Goremykina2019, Morningstar2020}.

\subsection{Strong coupling (region II)}
\label{strongCoupling}
\begin{figure}
  \centerline{\includegraphics[width=0.99\columnwidth, trim={0 0 0 0}, clip]{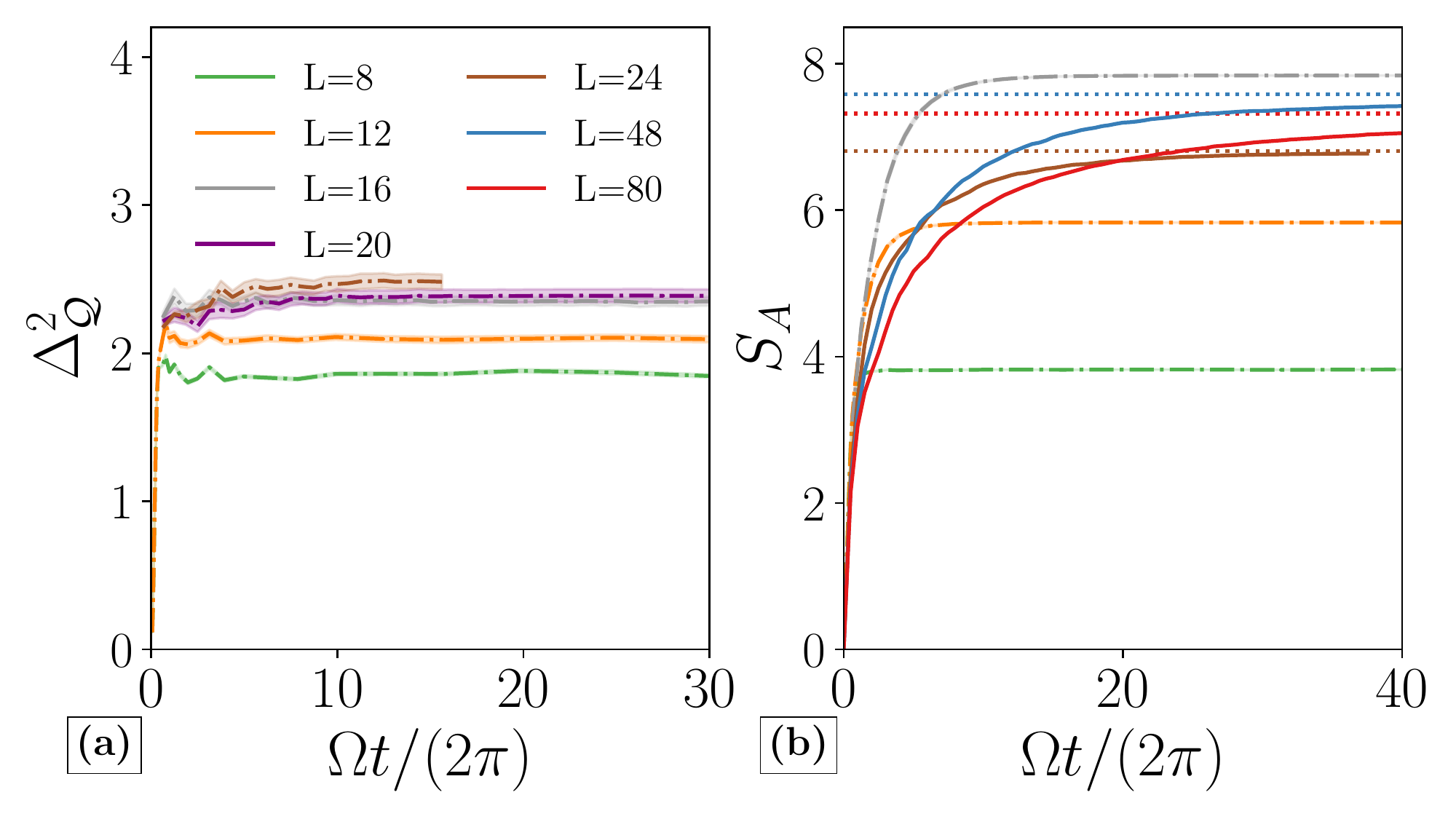}}
  \caption{Dynamics for strong coupling, $\gamma  = 0.7/\sqrt{L/12}$. \textbf{(a)} Qudit variance $\Delta^2_{\mathcal{Q}}$ and \textbf{(b)} bipartite entanglement entropy $S_A$. Results from ML-MCTDH are not included for $\Delta^2_{\mathcal{Q}}$ as they are not converged. For system sizes where the dynamics can be computed exactly (dot-dashed lines), $S_A$ saturates the Page bound $S_A \sim L/2$. The curves from ML-MCTDH (solid lines), corresponding to $L\geq 24$, saturate the bound set by the number of single-particle functions in the second layer, $\log_2{\chi_2}$ (dotted lines).}
  \label{fig:strongCoupling}
\end{figure}

In the strong coupling regime (Fig.\ \ref{fig:strongCoupling}), our phase diagram suggests that the system lies deep within the thermalizing phase for our available system sizes due to strong delocalizing interactions between the spins induced by the central qudit. Our data is consistent with this expectation, but we note two effects. First, despite the fact that spins thermalize, we observe that the asymptotic distribution of qudit occupations is nonuniform, similar to the athermal qudit regime found in \cite{Ng2019}. In section \ref{sec:randMat} of \cite{Note2} we introduce a phenomenological picture to explain this based on random matrix theory, suggesting that it is rare for the qudit to make transitions between widely separated states. 

Second, we note that due to the ergodic character of the dynamics in the strong-coupling thermalizing regime, the accurate treatment of the dynamics represents a significant challenge for the ML-MCTDH approach and cannot be converged for longer times \cite{Westermann2012}.
This well known limitation of the ML-MCTDH method and other tensor network approaches is due to the following reason. Within the ML-MCTDH approach, the wave function of the system is represented in each layer by sums of Hartree products, the total number of which is determined by the number $N_n$ of SPFs employed in a given layer $n$ for each degree of freedom. For example, in the binary tree depicted in Fig.\ \ref{fig:mctdh_tree},  $N_2$ SPFs are used in the second layer to represent each of the two parts of the spin chain resulting in $(N_2)^2$ Hartree products that represent the spin system in the second layer. As a consequence, the entanglement entropy between the different constituents of the system is bounded by $\log N_2$. However, for ergodic systems, the entanglement entropy is extensive, and thus, starting from an uncorrelated state, the entanglement entropy grows and eventually exceeds the limit of $\log N_2$. This implies that, for longer times, the wave function of the system cannot be represented accurately. The application of the ML-MCTDH formalism in the ergodic phase is thus restricted to short times. Therefore, the results for the qudit variance depicted in Fig.\ \ref{fig:strongCoupling} have been obtained by exact diagonalization and the kernel polynomial method.

Despite this limitation of ML-MCTDH, we are still able to find signatures of thermalization by examining the dependence of the dynamics on $N_2$.  In the right panel of Fig.\ \ref{fig:strongCoupling}, we see that the bipartite entanglement entropy is upper bounded by $\log_2 N_2$, corresponding to a maximal entropy state within our variational ansatz. We observe similarly strong dependence of $q_{\text{EA}}$ when increasing $N_2$, which drifts towards zero to indicate paramagnetic behavior in the spin chain. Other observables, such as the populations of the qudit levels cannot be converged, implying that information about the ergodic state is present, but limited.

\section{Results for unscaled coupling}
\label{sec:unscaledGamma}

The dynamics of our system with scaled coupling, $\gamma \sim 1/\sqrt{L}$, is perhaps most interesting because it gives a well-defined thermodynamic limit for the qudit. However, it is also important to understand the dynamics when the coupling is held fixed instead of being scaled by system size, corresponding to the dashed horizontal line in Fig.\ \ref{fig:phaseDiagram}. Fixing the coupling strength may be easier to implement experimentally, for example in cavity QED where the coupling is governed by the position-dependent electric field strength. Doing so, however, means that we can no longer easily separate dynamics occurring on different timescales as in the previous section. Furthermore, we predict that in the thermodynamic limit this will eventually result in thermalization, as the long-range interactions induced by the central qudit will eventually dominate at large enough times and system sizes. We choose the fixed value $\gamma=0.10607$, which precisely matches the intermediate scaled coupling for our largest system size, and therefore lives within the predicted MBL phase for all accessible system sizes.

\begin{figure}[H]
  \centerline{\includegraphics[width=0.99\columnwidth, trim={0 0 0 0}, clip]{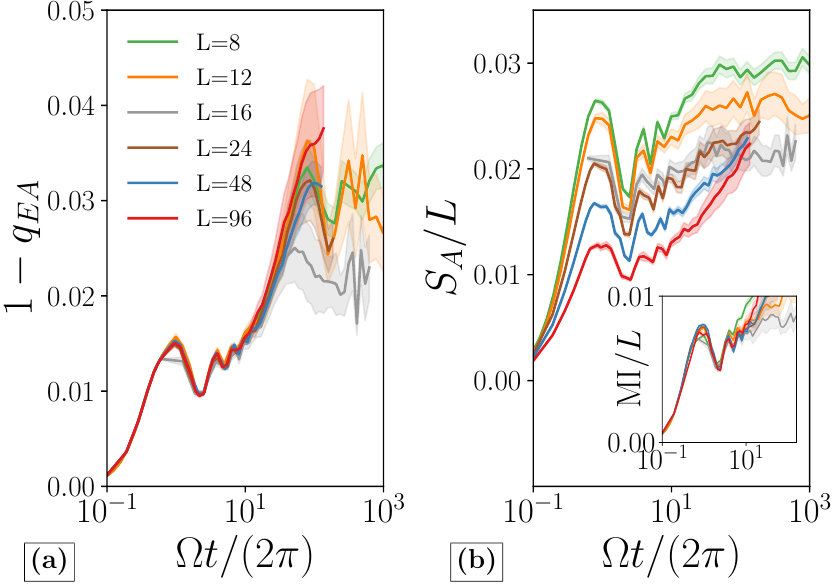}}
  \caption{Dynamics with unscaled coupling, $\gamma=0.10607$. On the reachable timescales, the spin glass order parameter $q_{\text{EA}}$ \textbf{(a)} does not show significant system size dependence. However, the bipartite entanglement entropy \textbf{(b)}, in addition to being subextensive -- $S_A \propto L^\alpha$, $0 < \alpha < 1$ -- at short times, already shows qualitatively different behavior at intermediate time for $L \gtrsim 24$. \textbf{(inset)} Mutual information $MI \equiv I(A,B)$ between two contiguous halves $A$ and $B$ of the spin chain (see main text).}
  \label{fig:unscaled}
\end{figure}

The dynamics with fixed coupling, shown in Fig.\ \ref{fig:unscaled}, looks similar to the data at intermediate scaled coupling but without as clear a separation of time scales or data collapse. We note that it is harder to detect the sort of logarithmically slow delocalization as seen in $q_{\text{EA}}$ in Fig.\ \ref{fig:intTime}f (cf.\ Fig.\ \ref{fig:unscaled}a). The story is the same with the qudit variance, which is similar to $1-q_{\text{EA}}$ when divided by $L$. However, with the fixed coupling, the entanglement entropy reflects a qualitative change in behavior at large enough system sizes. 

We see that at times $\frac{\Omega}{2\pi}t\sim 6\times 10^1$ in the localized phase (Fig.\ \ref{fig:unscaled}), the smaller system sizes $L\leq 16$ establish a subextensive amount of entanglement entropy. Numerics from ML-MCTDH seem to counter this trend, with $S_A$ continuing to grow slowly beyond this timescale. The rate of this growth increases with $L$, which is consistent with it arising from stronger effective all-to-all interactions described by the high-frequency expansion (Eq.\ \ref{eq:HFE}). It also appears to show strong system size dependence at short times, where a subextensive amount of entanglement is established. This should be contrasted with models of MBL without central coupling, in which the short time behavior is system size independent.

As a final note, we point out that the entanglement entropy at fixed $\gamma$ is subextensive, such that $S_A/L$ appears to be trending towards zero with increasing system size. This should be contrasted with mutual information between the two halves of the spin chain,
\begin{equation*}
I(A,B)=S(A) + S(B) - S(A \cup B),
\end{equation*}
which is extensive. In Ref.\ \cite{Ng2019}, mutual information was used as a proxy for entanglement between the two halves of the spin chain, as it nominally removes ``unimportant'' entanglement with the central qudit. However, since entanglement must be subextensive -- indeed, system size independent -- in the MBL phase, our data indicate that entanglement entropy is a better metric than mutual information for capturing this. Our initial expectation was that mutual information would become subextensive at larger system size, but the results obtained with the ML-MCTDH method rule out that possibility.

\section{Conclusions}
In this paper we have studied the dynamical behavior of a qudit coupled to a disordered, interacting bath of up to $L=96$ spins-$1/2$, which altogether can exhibit localization at strong disorder. Using a combination of exact propagation methods and the tensor network-based ML-MCTDH approach, we find evidence of qualitatively different dynamical signatures in local observables such as the spin glass order of the bath and the qudit variance, consistent with a rough phase diagram (Fig.\ \ref{fig:phaseDiagram}). Most notably, we find hints of logarithmically slow decay of localization near the onset of all-to-all interactions in the bath. This behavior was found to occur after timescales $t\sim O(1/\gamma^2)$ where $\gamma$ is the qudit-spin bath coupling. 

The behavior of the qudit observed here is, we believe, not specific to this model. Our conclusions should apply equally well to the cases of a cavity photon with rescaled raising/lowering operators $a^\dagger \to (N_0)^{-1/2} a^\dagger$ or central spin-$S$ systems with operators rescaled as $\hat{S} \to (S(S-1))^{-1/2} \hat{S}$. The feature of these systems is that the fundamental commutation relation between the raising and lowering operators vanishes in the limit of large $S$ or large $N_0$. This fact allows for exact cancellation between processes that raise or lower the qudit state. However, this mechanism only serves to protect localization for sufficiently large ``magnetic field'' $\Omega$; it is unclear how these systems interpolate between the $\Omega=0$ limit and the $\Omega > |g|, |h_i|, \ldots$ limit. We note additionally that the limitations of ML-MCTDH for these types of centrally coupled systems with many-body interacting baths in the strong coupling regime requires more clarification. Such clarifications may be necessary to extend the effectiveness of the method into the thermalizing regime on the left side of the phase diagram \ref{fig:phaseDiagram}, which remains numerically inaccessible and thus poorly understood.

\begin{center}
    \textbf{Acknowledgements}
\end{center}
This work was performed with support from the National Science Foundation through award number DMR-1945529 (MHK), the Welch Foundation through award number AT-2036-20200401 (MHK), and the German Research Foundation (DFG) through IRTG 2079.
This research used resources of the National Energy Research Scientific Computing Center, a U.S. Department of Energy Office of Science User Facility operated under Contract No. DE-AC02-05CH11231. Furthermore, support by the state of Baden-Württemberg through bwHPC and the DFG through grant no.\ INST 40/575-1 FUGG (JUSTUS 2 cluster) is gratefully acknowledged.

\bibliographystyle{apsrev4-1}
\bibliography{references}

\onecolumngrid
\newpage

\appendix
\begin{center}
\Large
    Supplemental Material for ``Localization dynamics in a centrally coupled system''
\end{center}

\section{Linear response property in the thermodynamics limit}
In the limit $L\to \infty$ the spin chain can be seen as a macroscopic environment allowing for a different viewpoint on the dynamics of the qudit. Environments consisting of independent (i.e. noninteracting) degrees of freedom and appropriate scaling of the coupling $\gamma$ fulfil linear response theory, i.e. the influence of the environment on the qudit is fully characterized by the force autocorrelation function of the bath allowing for the description of the influence of the environment in terms of a bath of harmonic oscillators with an effective spectral density.\cite{Makri1999} This idea was used, for example, to describe the dynamics of a two-level system coupled to a spin bath consisting of independent spins\cite{Shao1998, Makri1999} and to a bath of anharmonic vibrational degrees of freedom\cite{Wang2007}. For non-interacting environments this was proven by showing that all but the leading order term in the cumulant expansion of the influence functional vanish in the thermodynamic limit. Despite the fact that the spin chain considered in our work consists of interacting spins, one can show that the linear response property also holds for this model, and thus, the influence of the spin chain on the qudit can be described by a bath of harmonic oscillators with an effective spectral density. 

To this end, we follow the derivation given in Ref.\,[\onlinecite{Makri1999}] and extend it to the interacting spin chain considered here. We assume that initially there are no correlations between different spins and the spin chain is in an eigenstate of all $\lbrace\sigma_i^z\rbrace$. In order to evaluate the different terms in the cumulant expansion\cite{Makri1999} the time dependent operator
\begin{align}
\hat{f}(t) &= - \gamma \sum_{i=1}^L \underbrace{{\rm e}^{i H_0 t} \sigma_i^x {\rm e}^{-i H_0 t}}_{\sigma_i^x(t)}
\end{align}
is required, which corresponds to the bath part of the system-bath interaction and describes the force exerted on the system due to its interaction with the environment. Here, $H_0$ is the Hamiltonian of the isolated spin chain. The time evolved operators $\sigma_i^x(t) = \sigma_i^+(t) + \sigma_i^-(t)$ can be calculated analytically yielding
\begin{align}
\sigma_i^+(t) &= {\rm e}^{2ih\xi_it} \frac{1}{2} \bigg[ [\mathds{1}- \sigma_{i-1}^z\sigma_{i+1}^z] + [\mathds{1} + \sigma_{i-1}^z\sigma_{i+1}^z]\cos(4gt)  + i [\sigma_{i-1}^z+\sigma_{i+1}^z]\sin(4gt) \bigg] \sigma_i^+ \nonumber \\
&= {\rm e}^{2ih\xi_it} \hat{\varphi}_i(t) \sigma_i^+ \label{eq:tdp-operator},
\end{align}
and $\sigma_i^-(t) = (\sigma_i^+(t))^\dagger$. Here $\sigma_i^+$ and $\sigma_i^-$ are the spin raising and lowering operators, respectively. Using this equation, one can show that 
\begin{align}
\big[\sigma_i^x(t), \sigma_j^x(t^\prime) \big]& = 0 ~~~~ \forall~ i,j ~{\rm with} ~|i-j| \geq 2,
\end{align}
This follows from the fact that $\hat{\varphi}_i(t)$ only involves spin operators on sites $i-1$ and $i+1$.
 
All terms in the cumulant expansion of the influence functional can be expressed\cite{Makri1999} in terms of $N$-time correlation functions defined as
\begin{align}
C^{(N)}(t_1,..., t_N) &= \braket{\hat{f}(t_1)...\hat{f}(t_N)}_0,
\end{align}	
where $\braket{...}_0$ denotes the expectation value with respect to the initial state of the environment. Initially, the spin chain is in an eigenstate of all $\lbrace\sigma_i^z\rbrace$, and thus, one finds that
\begin{align}
C^{(1)} = 0,
\end{align}
Consequently, the first term, and by extension all odd order terms in the expansion, vanishes. The second order term can be expressed in terms of the two-point correlation function
\begin{align}
C^{(2)}(t_1, t_2) &= \gamma^2 \sum_i \sum_j \braket{\sigma_i^x(t_1)\sigma_j^x(t_2)}_0.
\end{align}
It is straightforward to check that the expectation value is zero for all $i \neq j$, and thus, the two-time correlation function reduces to
\begin{align}
C^{(2)}(t_1, t_2) &=  \gamma^2 \sum_{i=1}^L \braket{\sigma_i^x(t_1)\sigma_i^x(t_2)}_0.
\end{align}
Since $\braket{\sigma_i^x(t_1)\sigma_i^x(t_2)}_0$ is finite but not zero in general, the sum diverges in the thermodynamic limit $L\to \infty$ unless $\gamma\sim \nicefrac{1}{\sqrt{L}}$, which we will assume in the following by setting $\gamma = \gamma_0\sqrt{\nicefrac{L_0}{L}}$. The next non-vanishing term in the cumulant expansion is the fourth term which involves the fourth order correlation function
\begin{align}
C^{(4)}(t_1, t_2, t_3, t_4) &= \gamma^4 \sum_{i,j,k,l} \braket{\sigma_i^x(t_1)\sigma_j^x(t_2)\sigma_k^x(t_3)\sigma_l^x(t_4)}_0.
\end{align}
Since $\hat{\varphi}_i(t)$ does not change the initial state of the spin chain, there can only be up to two different indices in the expectation value. Thus, the four-time correlation function can be written as
\begin{align}
C^{(4)}(t_1, t_2, t_3, t_4) &= \gamma^4 \sum_{i,j} \braket{\sigma_i^x(t_1)\sigma_i^x(t_2)\sigma_j^x(t_3)\sigma_j^x(t_4)}_0 \nonumber \\
&+\gamma^4 \sum\limits_{\substack{i,j \\i \neq j}} \braket{\sigma_i^x(t_1)\sigma_j^x(t_2)\sigma_i^x(t_3)\sigma_j^x(t_4)}_0 \nonumber\\
&+\gamma^4 \sum\limits_{\substack{i,j \\i \neq j}} \braket{\sigma_i^x(t_1)\sigma_j^x(t_2)\sigma_j^x(t_3)\sigma_i^x(t_4)}_0.
\end{align}
In the following we will discuss only the first term on the right hand side. The other two terms can be treated equivalently. The double sum can be decomposed as
\begin{align}
\gamma^4 \sum_{i,j} \braket{\sigma_i^x(t_1)\sigma_i^x(t_2)\sigma_j^x(t_3)\sigma_j^x(t_4)}_0 &= \gamma^4 \sum_{i} \braket{\sigma_i^x(t_1)\sigma_i^x(t_2)\sigma_i^x(t_3)\sigma_i^x(t_4)}_0 \nonumber\\
&+ \gamma^4 \sum_{i} \braket{\sigma_i^x(t_1)\sigma_i^x(t_2)\sigma_{i+1}^x(t_3)\sigma_{i+1}^x(t_4)}_0\nonumber \\
&+ \gamma^4 \sum_{i} \braket{\sigma_{i+1}^x(t_1)\sigma_{i+1}^x(t_2)\sigma_i^x(t_3)\sigma_i^x(t_4)}_0 \nonumber\\
&+ \gamma^4 \sum\limits_{\substack{i,j \\ |i-j|\geq2}}^L \braket{\sigma_i^x(t_1)\sigma_i^x(t_2)\sigma_j^x(t_3)\sigma_j^x(t_4)}_0.
\end{align}
The expectation values in the first three sums on the right hand side are bounded by a constant. If the coupling $\gamma$ is scaled as $\gamma = \gamma_0\sqrt{\nicefrac{L_0}{L}}$, these terms vanish as $L\to \infty$, and thus, we conclude that
\begin{align}
\gamma^4 \sum_{i,j}^L \braket{\sigma_i^x(t_1)\sigma_i^x(t_2)\sigma_j^x(t_3)\sigma_j^x(t_4)}_0 &= \gamma^4 \sum\limits_{\substack{i,j \\ |i-j|\geq2}}^L \braket{\sigma_i^x(t_1)\sigma_i^x(t_2)\sigma_j^x(t_3)\sigma_j^x(t_4)}_0  + \mathcal{O}\bigg(\frac{1}{L}\bigg).
\end{align}
Because the operators in the expectation value act on different Hilbert spaces and the initial state factorizes the expectation values can be factorized as
\begin{align}
\gamma^4 \sum\limits_{\substack{i,j \\ |i-j|\geq2}}^L \braket{\sigma_i^x(t_1)\sigma_i^x(t_2)\sigma_j^x(t_3)\sigma_j^x(t_4)}_0 &= \gamma^4 \sum\limits_{\substack{i,j \\ |i-j|\geq2}}^L \braket{\sigma_i^x(t_1)\sigma_i^x(t_2)}_0 \braket{\sigma_j^x(t_3)\sigma_j^x(t_4)}_0  
\end{align}
Adding the terms for $i=j$ and $|i-j|=1$ to the double sum on the right hand side gives an error of $\mathcal{O}(\nicefrac{1}{L})$, and thus, one can write
\begin{align}
\gamma^4 \sum_{i,j} \braket{\sigma_i^x(t_1)\sigma_i^x(t_2)\sigma_j^x(t_3)\sigma_j^x(t_4)}_0 &= \gamma^4 \sum_{i,j} \braket{\sigma_i^x(t_1)\sigma_i^x(t_2)}_0 \braket{\sigma_j^x(t_3)\sigma_j^x(t_4)}_0 + \mathcal{O}\bigg(\frac{1}{L}\bigg) \\
&= C^{(2)}(t_1, t_2) C^{(2)}(t_3, t_4) + \mathcal{O}\bigg(\frac{1}{L}\bigg),
\end{align}
where we have identified the two-time correlation functions. With this, we finally conclude that
\begin{align}
\lim_{L\to \infty} C^{(4)}(t_1, t_2, t_3, t_4) = \lim_{L\to \infty} ~& C^{(2)}(t_1, t_2) C^{(2)}(t_3, t_4)\nonumber \\
&+C^{(2)}(t_1, t_3) C^{(2)}(t_2, t_4) \nonumber  \\
&+C^{(2)}(t_1, t_4) C^{(2)}(t_2, t_3).
\end{align}
Using this one finds that the fourth order term in the cumulant expansion in [\onlinecite{Makri1999}] vanishes in the thermodynamic limit. In a similar way one can show that all higher order terms in the expansion vanish, proving that in the thermodynamic limit the influence functional is completely characterized by the force autocorrelation function. Thus, one can construct a bath of harmonic oscillators with an effective spectral density resulting in the same influence functional.

As a last step we show that for the choice of the 'super-Neel' state $\ket{E_{SN}} = \ket{\uparrow\uparrow\downarrow\downarrow...}$ as initial state the force autocorrelation function does not depend on the spin-spin interaction $g$. The force autocorrelation function is defined as
\begin{align}
\braket{\hat{f}(t_1)\hat{f}(t_2)} &= \gamma_L^2 \sum_{i} \braket{E_{SN}|\sigma_i^x(t_1)\sigma_i^x(t_2)|E_{SN}},
\end{align}
where the time-dependent operators are given in equation (\ref{eq:tdp-operator}). For the "super-Neel" state it follows that
\begin{align}
\sigma_{i-1}^z\sigma_{i+1}^z\ket{E_{SN}} &= -\ket{E_{SN}} \nonumber, \\
\big(\sigma_{i-1}^z+\sigma_{i+1}^z\big)\ket{E_{SN}} &= 0,
\end{align}
holds for all $i$ since the spins at site $i-1$ and $i+1$ are always antiparallel. Thus, the action of $\hat{\varphi}_i(t)$ is independent of the index $i$ and gives
\begin{align}
\hat{\varphi}_i(t) \ket{E_{SN}} = \ket{E_{SN}}.
\end{align}
Since the spin-spin interaction enters only via $\hat{\varphi}_i(t)$, we find that the force autocorrelation function, and consequently the parameters of the effective bath harmonic oscillators, are independent of the spin-spin interaction $g$. The corresponding effective spectral density can be calculated from the force autocorrelation\cite{Makri1999} yielding 
\begin{align}
J_{\rm eff}(\omega) &= \frac{\pi}{2} \frac{1}{4h} \chi_{[-2h, 2h]}(\omega),
\end{align}
where $\chi_{I}$ is the characteristic function of the interval $I$, i.e. $\chi_{I}(\omega)=1$ if $\omega\in I$ and 0 else. Thus, in the thermodynamics limit a bath of harmonic oscillators with this spectral density gives rise to the same qudit dynamics as the spin chain in the "Super-Neel" state.

\section{Structure of wavefunction}
\begin{figure}[H]
  \includegraphics[width=0.99\textwidth, trim={0 0 0 0}, clip]{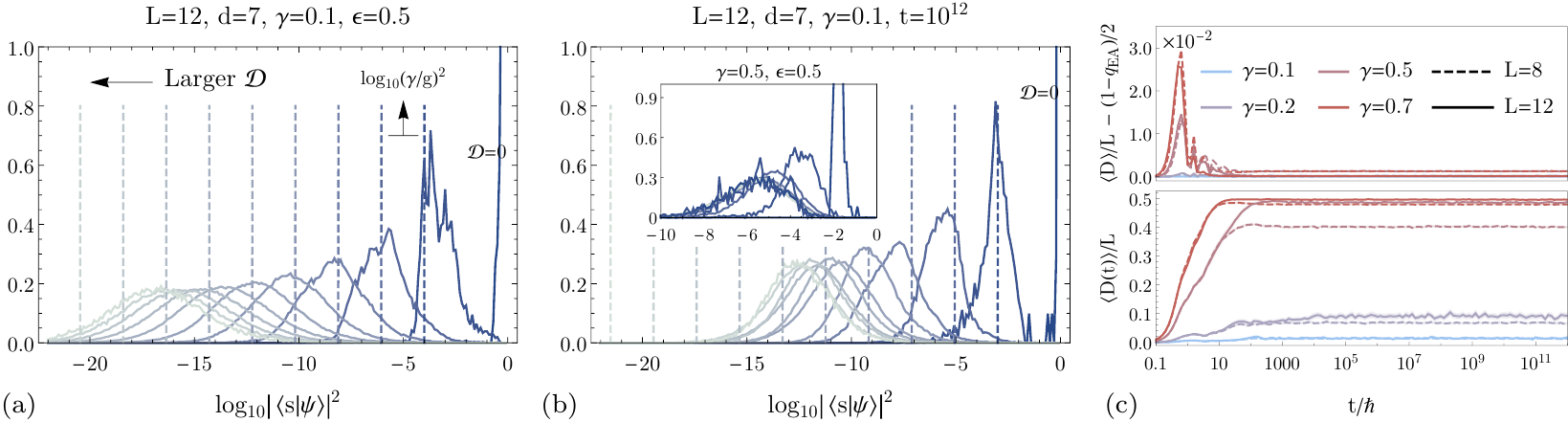}
  \caption{Distributions of wavefunction coefficients in the unperturbed basis $\{ |s\rangle \}$, parsed by the Hilbert space distance $\mathcal{D}$ separating $|s\rangle$ and a reference state $|s_{\text{ref}}\rangle$. The colors of the curves lighten for increasing $\mathcal{D}$, while the dashed vertical lines are guides to the eye for where the expected peak locations should the coefficients decay as $\propto (\gamma/g)^{2\mathcal{D}}$. \textbf{(a)} Distributions in the ten eigenstates closest to the middle of the energy spectrum, choosing $|s_{\text{ref}}\rangle$ to be the unperturbed state with largest weight. \textbf{(b)} Distribution of coefficients for the system at $t=10^{12}$, evolving from $|\psi(0)\rangle = |s_{\text{ref}}\rangle|(d+1)/2\rangle$ with $|s_{\text{ref}}\rangle$ being the super-Neel state. \textbf{(c)} Average Hilbert space distance $\lrangle{D(t)}$ from the super-Neel state (bottom) and in comparison with the spin glass order parameter $q_{EA}$ (top), for $\gamma = 0.1, 0.2, 0.5, 0.7$ and for $L=8\text{(dashed)}, 12\text{(solid)}$.}
  \label{fig:wavefunction}
\end{figure}

In addition to modifying the entanglement dynamics at short times, the star-like geometry of this system (depicted in the inset of Figure \ref{fig:mctdh_tree}) should render the concept of locality meaningless. Indeed, from the point of view of operator dynamics, operators for the qudit should immediately spread to $O(L)$ sites after $O(1)$ time\cite{Lucas2019}. Instead we propose to analyze the structure of eigenstates in the Hilbert space of the uncoupled ($\gamma=0$) Hamiltonian. In the spirit of the current understanding of MBL, where eigenstates are weak deformations of the unperturbed system, we parse wavefunctions in the product basis $|s\rangle$ of spins and qudit states -- $|\vect{z}\rangle\otimes|n\rangle$ with $z_i \in \{\uparrow, \downarrow\}, \forall i=1,\ldots,L$ and $n=1,\ldots,d$. We quantify ``deformations'' through a notion of Hilbert space distance as 
\begin{align}
\label{eq:hamDis}
\mathcal{D}(|\vect{z},n\rangle, |\vect{z}',n'\rangle) = \max \left( \mathcal{D}_H(\vect{z}, \vect{z}'), |n - n'| \right),
\end{align}
with $\mathcal{D}_H$ being the Hamming distance between the bitstrings $\vect{z}$ and $\vect{z}'$ (here, up-spins are ``1'' and down-spins are ``0''). Intuitively, this measures the minimum number of times the transverse perturbation must be applied to connect two states in the Hilbert space. For eigenstates, the origin ($\equiv |\psi_{\text{ref}}\rangle$) is taken to be the product state with the largest weight while for time evolution, the origin is taken to be the initial state before the quench. Other product states can then be grouped according to their distance $\mathcal{D}$ from $|\psi_{\text{ref}}\rangle$. For each $\mathcal{D}$, we calculate the distribution of expansion coefficients $|\lrangle{s | \psi}|^2$ over disorder realizations and equidistant product states $\{|s\rangle \mid \mathcal{D}(|\psi\rangle, |s\rangle) = x\}$. In the nonergodic phase, these coefficients are suppressed by a factor of $(\gamma/g)^2$ as $\mathcal{D}$ increases (Figure \ref{fig:wavefunction}(a,b)). One can then regard the wavefunction as being exponentially localized in Hilbert space. This can be observed also in non-centrally coupled models of MBL, such as the disordered Ising chain with next-nearest neighbor interactions \cite{Kjall2014}. In contrast, there is no such Hilbert space localization at large enough $\gamma$ in the ergodic phase (inset of Figure \hyperref[fig:wavefunction]{5b}). This is corroborated by the average Hilbert space distance $\lrangle{\mathcal{D}(t)}$, as measured from $|\psi_{\text{ref}}\rangle$. This quantity has been noted by Hauke and Heyl \cite{Hauke2015} to saturate to $L/2$ if the system is (possibly) ergodic, and is consistent with our numerics (Figure \hyperref[fig:wavefunction]{5c}). We additionally find that the experimental accessibility $\lrangle{\mathcal{D}(t)}$ (noted by \cite{Hauke2015}) approximately holds (see top panel of Figure \ref{fig:wavefunction}c), i.e.\ $\lrangle{\mathcal{D}(t)} \geq L (1-q_{EA})/2$ where the inequality is due to the definition of $\mathcal{D}$ we have chosen. 

In the top panel of Figure \ref{fig:wavefunction}c, the periods where $\mathcal{D}/L > (1-q_{EA})/2$ are due to the fast increase and saturation of qudit variance on the timescale of $\sim 1/\gamma$, versus the slower decay of $q_{EA}$. The latter proceeds on slower timescales through a combination of qudit-spin flip transitions and the Ising interaction $g \spin{z}{i}\spin{z}{i+1}$. This interpretation of the spin glass order parameter bring new meaning to the results from ML-MCTDH. At least up to intermediate times, the localization length of the wavefunction in Hilbert space is stable up to $L=96$ when $\gamma \approx 0.106$.

That the system is localized in Hilbert space may be useful in improving the performance of ML-MCTDH in this regime. Currently, ML-MCTDH reduces the size of the full Hilbert space by restricting the dynamics on to a subspace created by uniformly random vectors. For very large systems, these randomly drawn vectors will be heavily weighted towards states farther away from the initial/reference state, since their numbers grow combinatorially quickly. 

\section{Trivial limit and convergence of qudit variance}
\label{sec:appA}
\begin{figure}[h]
  \centerline{\includegraphics[width=0.95\textwidth, trim={0 0 0 0}, clip]{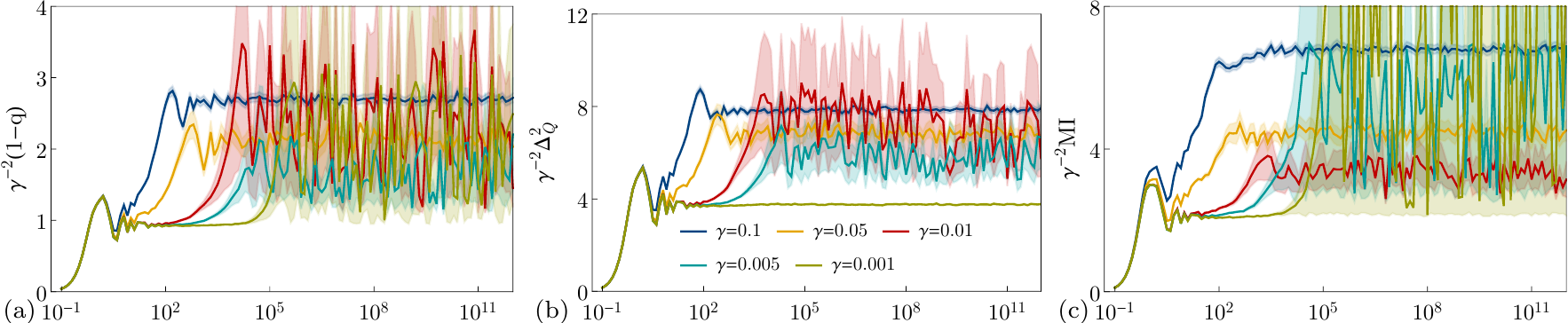}}
  \caption{Dynamics from the ``super-Neel'' state for $L=8$ across different $\gamma$ in the localized phase. Rescaling the \textbf{(a)} spin glass order parameter, \textbf{(b)} qudit variance, and \textbf{(c)} mutual information by the coupling collapses the dynamics at short to intermediate times. The asymptotic behavior in each case should be interpreted as defining a trivial limit analogous to Anderson localization vis-a-vis MBL. }
  \label{fig:scaledL8}
\end{figure}

The quench setup we examine in this paper -- in which we prepare the full system in an eigenstate of the $\gamma = 0$ Hamiltonian -- results in certain behaviors in the $\gamma\to 0$ limit which we will explore in this section. The reason for the existence of a well-defined limit in the dynamics is due to the off-diagonal coupling being the \textit{only} generator of dynamics in both the spins and the qudit at short times. 
We demonstrate this limit by plotting observables rescaled by $\gamma^{-2}$ in Figure \ref{fig:scaledL8}. This scaling at small $\gamma$ converges the dynamics at short times up to $t\leq 10^2$. At small enough coupling, a plateau begins to appear at $t \sim 10^2$. We see hints of this in the MCTDH data for mutual information (Figure \ref{fig:intTime}c), for example, with the establishment of a plateau for $L=48$ and $80$ at times $10 \leq t \lesssim 10^2$. This behavior contrasts with the immediate increase of MI for smaller system sizes at $t\sim 10$.

We take advantage of the vanishing coupling, which is scaled to zero with system size, to systematically reconstruct the dynamics using the method of multiple scales. To that end, we solve for the time evolution operator, artificially introducing new `independent' timescales $t$, $t' \equiv \gamma t$, $t'' \equiv \gamma^2 t$, $\ldots$, which allow for control over secular terms growing unboundedly as $\sim t$. Formally, the time development operator is expanded as $U(t) \equiv U_0(t, t', t'', \ldots) + \gamma U_1(t, t', t'', \ldots) + \gamma^2 U_2(t, t', t'', \ldots) + \ldots$ and we shall solve for the full evolution order-by-order. We note that we have made an important assumption that the only timescales of interest are $\sim \gamma^{-n}$. However, we are not interested in describing the dynamics for all $t$ at arbitrary $\gamma$ and $L$, but only up to the $t \sim \gamma^{-1}$ as $\gamma\to 0$.

The propagator evolves according to
\[ \frac{d}{dt} U(t) = -i (H_0 + \Omega \tau^z + \gamma H_1 \tau_x) U(t), \] where we define $\displaystyle \tau^z = \sum_{n=1}^d n|n\rangle\langle n|$, $\displaystyle \tau^+ = \sum_{n=1}^{d-1} |n+ 1\rangle\langle n|$, and $\tau^- = (\tau^+)^\dagger$. From these we construct $\tau^x = \tau^+ + \tau^-$ and $\tau^y = -i \tau^+ + i \tau^-$. With the addition of the new timescales, the time derivative now becomes
\[ \frac{d}{dt} U = \left( \frac{\partial}{\partial t} U_0 \right)
  + \gamma \left( \frac{\partial}{\partial t'} U_0 + \frac{\partial}{\partial t} U_1 \right)
  + \gamma^2 \left( \frac{\partial}{\partial t''} U_0 + \frac{\partial}{\partial t'} U_1 + \frac{\partial}{\partial t} U_2 \right)
  + \ldots \]

At the zeroth order, the equation of motion and its solution are
\begin{align*}
\frac{\partial}{\partial t} U_0 = -i (H_0 + \Omega \tau^z) U_0 && \Longrightarrow && U_0 = e^{-i (H_0 + \Omega \tau^z) t} U^{\text{int}}_0(t', t'', \ldots) \text{ such that } U^{\text{int}}_0(0, 0, \ldots) = 1.
\end{align*}

At first order,
\begin{align}
  \frac{\partial}{\partial t'} U_0 + \frac{\partial}{\partial t} U_1 &= -i (H_0 + \Omega \tau^z) U_1 -i H_1 \tau^x U_0 \nonumber \\
  \frac{\partial}{\partial t'} U^{\text{int}}_0 + \frac{\partial}{\partial t} U^{\text{int}}_1 &= -i e^{i (H_0 + \Omega \tau^z) t}  H_1 \tau^x e^{-i (H_0 + \Omega \tau^z) t} V_0, \label{eq:appA_firstOrder}
\end{align}
where we let $U_1 = e^{-iH_0 t} U^{\text{int}}_1$. Note at this point that the first term on the LHS is independent of $t$. Its contribution to $U^{\text{int}}_1$ would be proportional to $t$ and thus secular. Should there also exist secular terms on the RHS (i.e., independent of $t$), $U^{\text{int}}_0$ should be chosen to offset it. Otherwise, it must be independent of $t'$, i.e.\  $U^{\text{int}}_0 \equiv V_0(t'', \ldots)$. If these secular terms do not exist at all orders, then the multiple scales result would be completely equivalent to the Dyson series in the interaction picture. In our disordered system, we must be careful of secularity and near-secularity. The former, in which two states linked by the perturbation are exactly degenerate, occurs with zero probability since the local field must have a value such that $\Omega = \pm 2(h_i + g(\spin{z}{i-1}+\spin{z}{i+1}))$. More likely is the scenario of near-degeneracies, which at this order can lead to $U^{\text{int}}_1$ growing arbitrarily large after an arbitrarily long time. Such terms make the expansion of $U$ uncontrolled at long times. We can, however, absorb near-secular behavior into $U^{\text{int}}_0$.

Define
\[ A(t') = \sum_{\substack{|a\rangle, |b\rangle \\ |E_a - E_b| < 1}} \lrangle{a | H_1 \tau^x | b} \exp\left(i \frac{\Delta E_{ab}}{\gamma} t'\right) |a\rangle\langle b|. \] To regulate the secular part of (\ref{eq:appA_firstOrder}), we must have $\frac{\partial}{\partial t'} U^{\text{int}}_0 = -i A(t') U^{\text{int}}_0$. This makes \[ U^{\text{int}}_0(t', t'',\ldots) = \exp\left(-i \int\limits_0^{t'} A(\tau) d\tau \right) V_0(t'', \ldots). \] The unknown function $V_0$ will be solved for at higher orders. The argument of the exponential should always be complex, since $A(t')$ is Hermitian. Thus these resonant terms will not cause $U^{\text{int}}_0$ to have unbounded norm, and the perturbative expansion for $U$ remains valid. However, our ability to regulate the secularity in this way should not be taken as a statement on the dynamics being localized. Instead, it implies that more careful consideration of $U^{\text{int}}_0$ is necessary to understand if resonances are able to cause delocalization. At present, all we need is to examine if we can safely neglect $\exp(-i \int^{\gamma t}_0 A)$ if we scale $\gamma \to 0$ with the inverse system size. We can think of $A$ as roughly being the adjacency matrix for states in Hilbert space, where two states are connected by an edge if they are resonant. It is known that the eigenvalues of adjacency matrices are bounded above by the maximum number of edges connecting to a vertex in the graph. A naive upper bound for our system is $2L$, which is the number of states that can be reached by applying the perturbation on to the eigenstates of the unperturbed system. Should the spectrum of $A$ saturate this bound, the matrix exponential will contain time dependence going as $\exp(-i 2 \gamma t L)$ and scaling $\gamma \propto L^{-1/2}$ will not remove the correction factor in $U^{\text{int}}_0$. Regardless, this will not pose a problem in our model for the parameters and initial super-Neel state we have chosen.

Having found the lowest order approximation for $U^{\text{int}}_0$, we are now left with the nonsecular terms. These can be straightforwardly used to solve for $U^{\text{int}}_1$. Let $(H_1 \tau^x)_{\text{reg}}$ be the regular version of the perturbation $H_1 \tau^x$ with resonant matrix elements removed. The equation of motion becomes
\begin{align*}
  \frac{\partial}{\partial t}U^{\text{int}}_1 &= -i \, e^{i (H_0 + \Omega \tau^z) t}  (H_1 \tau^x)_{\text{reg}} e^{-i (H_0 + \Omega \tau^z) t} \, U^{\text{int}}_0(t', \ldots) \\
  \Longrightarrow \qquad U_1 &= -i \, e^{-i (H_0 + \Omega \tau^z) t} \left\{ i V_1(t', t'', \ldots) + \left( \int\limits_0^t e^{i (H_0 + \Omega \tau^z) \tau}  (H_1 \tau^x)_{\text{reg}} e^{-i (H_0 + \Omega \tau^z) \tau} d\tau \right) U^{\text{int}}_0(t', \ldots) \right\}
\end{align*} 

While we can in principle keep going to higher orders, we will stop here and discuss the dynamics with scaled qudit-spin chain coupling. The results we have obtained so far allow us to accurately describe time evolution up to a timescale $t \sim O(1/\gamma)$. Should we keep decreasing $\gamma$, then all the artificial times $t'$, $t''$, $\ldots$, will tend to zero without affecting the physical time $t$. Using the initial condition for the unknown functions can then give us closed form expressions for $U$. For example, we approximate
\[ U(t) \approx e^{-i (H_0 + \Omega \tau^z) t} + \gamma \, e^{-i (H_0 + \Omega \tau^z) t} \left(-i \int\limits_0^t e^{i (H_0 + \Omega \tau^z) \tau}  (H_1 \tau^x)_{\text{reg}} e^{-i (H_0 + \Omega \tau^z) \tau} d\tau \right). \] This is essentially what one finds in usual perturbation theory, except we now have more knowledge of its convergence properties.

We can gain some analytical understanding of the qudit variance and the spin glass order parameter from this approximation to the propagator.

Numerics show that the dynamics of qudit variance $\Delta^2_{\mathcal{Q}} = \lrangle{\tau(t)^2} - \lrangle{\tau(t)}^2$ comes mostly from the first term, as the second term is essentially constant. We calculate $\frac{d}{dt} \Delta^2_{\mathcal{Q}} \approx \lrangle{\acomm{\tau^z}{\frac{d}{dt}\tau^z}}$, where $\frac{d}{dt} \tau^z = \gamma H_1 \tau^y$. 
\begin{align*}
  \frac{d}{dt} \Delta^2_{\mathcal{Q}} &\approx \gamma \lrangle{\psi | \acomm{\tau^z(t)}{H_1(t) \tau^y(t)} |\psi} \\
  &= \gamma \cancelto{0}{\lrangle{\acomm{{\tau^z}^{(0)}}{ H^{(0)}_1 {\tau^y}^{(0)}}} } + \gamma^2 \lrangle{\acomm{{\tau^z}^{(1)}}{ H^{(0)}_1 {\tau^y}^{(0)}}} + \gamma^2 \lrangle{\acomm{{\tau^z}^{(0)}}{ \left( H^{(1)}_1 {\tau^y}^{(0)} + H^{(0)}_1 {\tau^y}^{(1)} \right)}}.
\end{align*}
The first term is zero since the operator is off-diagonal. The last term must also vanish since it is not invariant with respect to redefinition of $\tau^z$, e.g.\ changing its spectrum from $(0, \ldots, d-1)$ to $(1, \ldots, d)$ by adding a constant term to the Hamiltonian. Indeed one can check that it is exactly cancelled by the $- \lrangle{\tau^z}^2$ term we have neglected in our approximation of the qudit variance. A calculation of the remaining term shows
\begin{align}
\label{eq:appA_qdVarRate}
  \frac{d}{dt} \Delta^2_{\mathcal{Q}} &\approx 2 \gamma^2 \sum_i \sum_{\pm} \frac{\sin(t \Delta E^{\pm}_i)}{\Delta E^{\pm}_i} \lrangle{\psi^{\pm}_i \middle| \left( \spin{x}{i} \tau^x \right)_{\text{reg}} \middle| \psi},
\end{align}
consistent with usual perturbation theory. Upon disorder averaging, we see that
\begin{align*}
  \overline{\Delta^2_{\mathcal{Q}}} &\approx 2 \gamma^2 L \int\limits_0^t d\tau \, \overline{ \sum_{\pm} \frac{\sin(\tau \Delta E^{\pm}_i)}{\Delta E^{\pm}_i} \lrangle{\psi^{\pm}_i \middle| \left( \spin{x}{i} \tau^x \right)_{\text{reg}} \middle| \psi}}.
\end{align*}
Thus the qudit variance -- along with other qudit observables such as the population -- converge to a single curve upon scaling the coupling as $\gamma \propto 1/\sqrt{L}$. Convergence towards this expression should be expected up to time $t \sim O(\gamma^{-1}) \propto O(\sqrt{L})$. For dynamics from the super-Neel state as we study here, the energy difference with spin $i$ flipped is $\Delta E_i = \pm 2 h_i$. There are no resonances for our chosen values of $h_i \in [-1.3, 1.3]$ and $\Omega \approx 3.93$, so we have exactly
\begin{align*}
  \Delta^2_{\mathcal{Q}} &\approx 2 \gamma^2 \sum_i \sum_{\pm} \frac{1-\cos\left(t (2 h_i s_i \pm \Omega)\right)}{(2 h_i s_i \pm \Omega)^2} \\
  \frac{d}{dt}\overline{\Delta^2_{\mathcal{Q}}} &\approx \gamma^2 L \left( \frac{\operatorname{Si}[t (2h + \Omega)] + \operatorname{Si}[t (2h - \Omega)]}{h} \right),
\end{align*}
where $\operatorname{Si}(t)$ is the sine integral. For fixed $\gamma$ at large enough $L$, this expression will violate the bound on the qudit variance, $(d^2-1)/12$, when all the states of the qudit are equally populated. Thus higher order terms are necessary to prevent this unphysical outcome.

We can similarly look at the spin glass order parameter, and find that for the super-Neel state,
\[ \frac{d}{dt} (1-q) = 4 \gamma^2 \frac{1}{L} \sum_i \sum_{\pm} \frac{\sin(t \Delta E^{\pm}_i)}{\Delta E^{\pm}_i}. \]
This shows the surprising fact that the qudit variance (cf.\ \eqref{eq:appA_qdVarRate}) and spin glass order are linearly related to each other in this limit. This motivates the scaling we take in plotting the results in Figures \ref{fig:shortTime} and \ref{fig:intTime} in the main text.

\section{Ansatz for qudit populations in the thermalized phase}
\label{sec:randMat}
\begin{center}
  \begin{figure}[H]
\includegraphics[width=\columnwidth]{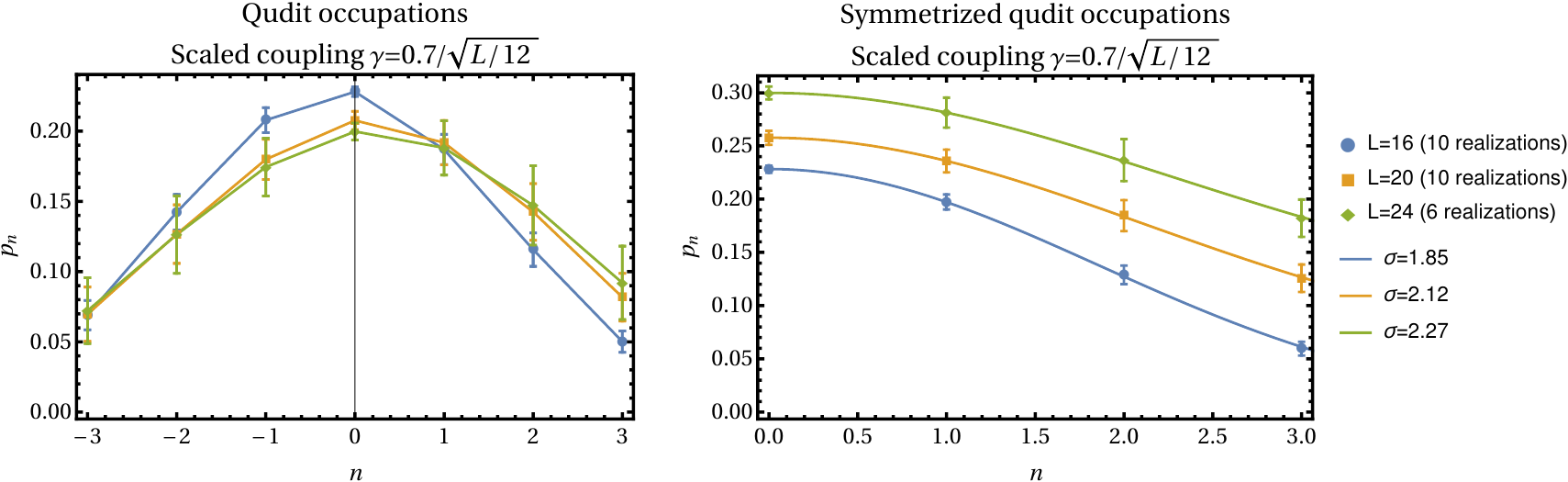}
  \caption{Occupations of the qudit at $t\sim 10^2 g^{-1}$. On the right, the occupations have been vertically spaced by $0.05$ for clarity.}
    \label{fig:occs}
  \end{figure}
\end{center}

We consider eigenstate thermalization in the sense that
\[ \lrangle{\psi | A |\psi } = Z^{-1} \Tr \left[ e^{-\beta H} A \right], \]
where $\beta$ is the temperature reproducing the same energy $\lrangle{\psi | H |\psi}$. When we time evolve from an initial state sitting at the middle of the many-body spectrum, $|\psi(0)\rangle = |\uparrow \uparrow \downarrow \downarrow \ldots\rangle \big| \frac{d-1}{2}\big\rangle$, we should consider infinite temperature averages, i.e.\
\[ \lrangle{\psi(t) | A |\psi(t) } = d^{-1} \Tr_{qd} \left( 2^{-L}  \Tr_S A \right). \] Therefore by setting $A = |n\rangle\langle n|$ for $n=(1-d)/2,\ldots,(d-1)/2$, we should expect a uniform occupation over all $d$ levels of the qudit. We do not observe this in the delocalized phase; instead, the occupations of the qudit seem to saturate to the distributions shown in Fig.\ \ref{fig:occs} (averaging over $\sim 10$ realizations of disorder).

On the left, we notice that the occupations appear to be asymmetric about the middle state. This is possibly due to the small number of disorder averages, but may also be affected by finite size effects or the rather special super-Neel initial state. As a proxy for the infinite sample limit, we symmetrize the occupations and note that a one parameter Gaussian ansatz, $p_n \propto \exp \left( -n^2/(2\sigma^2) \right)$, fits well. 

We propose a model to reproduce these observations based on the partial diagonalization achieved by invoking Floquet's theorem. By going into the rotating frame through the transformation $\exp \left( - i \Omega \hat{n} t \right)$, the Hamiltonian becomes time-periodic and we can factorize the time evolution in the ``lab frame'' as
\begin{align*}
  U(t) &= e^{-i \Omega \hat{n} t} e^{-i K_{\text{rot}}(t)} e^{-i H_{\text{rot}}^{\text{eff}} t} e^{i K_{\text{rot}}(0)}.
\end{align*} We had previously proven that eigenstates can be written in the form
\[ |E\rangle = e^{-i K_{\text{rot}}(0)} |\varepsilon_i\rangle |m\rangle, \qquad E = \varepsilon_i + m\Omega \text{ where } H_{\text{rot}}^{\text{eff}}|\varepsilon_i\rangle |m\rangle = \varepsilon_i |\varepsilon_i\rangle |m\rangle \text{ and } \hat{n}|m\rangle = m|m\rangle  \] under the assumptions that (1) the effective Hamiltonian $H_{\text{rot}}^{\text{eff}}$ commutes with $\hat{n}$ and (2) the operator \[ e^{-i \Omega \hat{n} t} e^{-i K_{\text{rot}}(t)} e^{i \Omega \hat{n} t} \] is an analytic function of time $t$. While explicit expressions for $K_{\text{rot}}$ and $H_{\text{rot}}^{\text{eff}}$ can be obtained using the high frequency expansion, the above expressions should hold even when the HFE does not converge, so long as the stated assumptions are satisfied. 

By Floquet's theorem, the kick operator $K_{\text{rot}}(t)$ must be time periodic with frequency $\Omega$. Hence it should be representable in a Fourier series in powers of $e^{-i \Omega m t}$. Because in the rotating frame, factors of $e^{\pm i \Omega t}$ are accompanied by the corresponding qudit raising/lowering operator $\tau^{\pm}$, we posit that terms in the Fourier series with $\exp(i \Omega m t)$ should induce transitions between qudit states separated by (signed distance) $m$. The kick operator should then decompose into
\[ K_{\text{rot}}(t) = \sum_{m=1}^{d-1} \sum_{n=-(d-1)/2}^{-m+(d-1)/2} e^{i \Omega m t} |n+m\rangle\langle n| \, B_n^{n+m} + \hc, \] where the operators $B^i_j$ act only on the spins. In the HFE, one sees that $B^i_j$ are imaginary and not necessarily Hermitian. We shall assume these properties still hold even when the HFE breaks down. 

We shall model the effect of the kick operator on only the qudit states by supposing that matrix elements of $B^i_j$ between two delocalized spin states are random numbers, with possible dependence on $i-j$. In short, we propose the replacement 
\[ \Tr_S e^{-i K_{\text{rot}}(0)} \rho e^{i K_{\text{rot}}(0)} \longrightarrow \overline{\exp(-i K) \rho_{qd} \exp(i K)}, \] 
where $K$ is a $d\times d$ Hermitian random matrix whose upper triangular part (excluding the diagonal) looks like
\[ \left( K \right)_{mn} = i g \exp \left( - \alpha \left( \frac{m-n}{\gamma/\Omega} \right)^2 \right) R_{mn}, \] for random $R_{mn} \sim \text{Normal}(\mu=0,\sigma^2=1)$ and $g,\alpha > 0$. The factor of $\gamma/\Omega$ was inserted so that $\exp(- i K)$ would be the unit matrix in the decoupled and infinite frequency limits. The average over all realizations of $K$ mimics the nonunitarity of the partial trace over the spins $S$. 

For example, we find good fits to the symmetrized occupations for the following values of the parameters, setting $g=1$:
\begin{center}
  \begin{tabular}{m{1.5in} m{3.9in}}
\begin{ruledtabular}
    \begin{tabular}[b]{@{}ccc@{}}
  $L$ & $\gamma$ & $\alpha$ \\ \hline
  16 & $0.7/\sqrt{16/12}$ & $1/105$ \\
  16 & $0.7$ & $1/83$ \\
  20 & $0.7/\sqrt{20/12}$ & $1/140$ \\
  24 & $0.7/\sqrt{24/12}$ & $1/175$
\end{tabular}
\end{ruledtabular} & 
                  \includegraphics[scale=0.7,trim={0 0pt 0 0pt}]{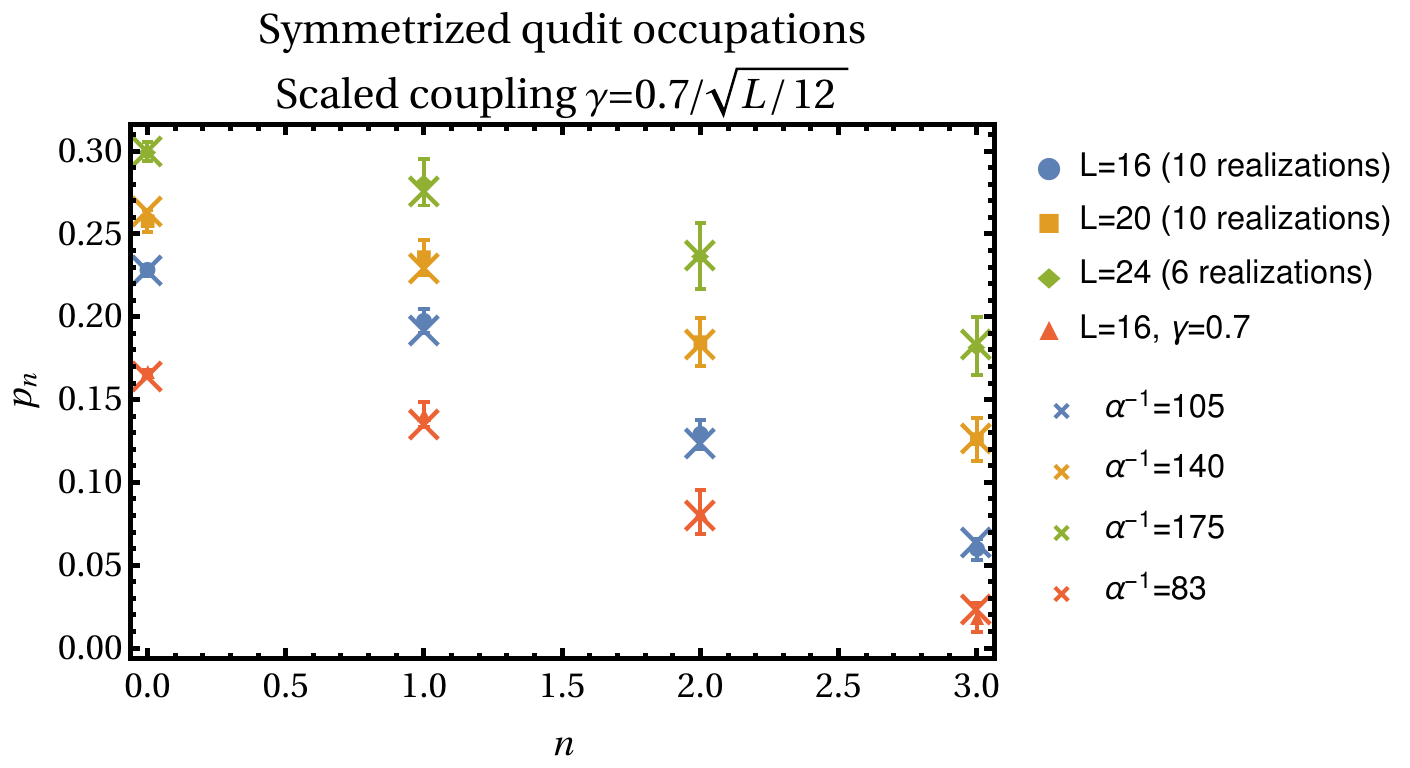}
                \end{tabular}
\end{center}

We conjecture that the correct form in the limit of large $L$ is
\[ \left( K \right)_{mn} = i \exp \left( - \frac{c}{L} \left( \frac{m-n}{\gamma/\Omega} \right)^2 \right) R_{mn}, \] where $c$ is a positive number of order 1.

\end{document}